\documentclass[letterpaper,twocolumn,10pt]{article}
\usepackage{usenix-2020-09}

\usepackage{tikz}
\usepackage{amsmath}
\usepackage{outlines}
\usepackage{framed}
\usepackage{array}
\usepackage[framemethod=TikZ]{mdframed} 
\usepackage{balance}
\usepackage{multirow, rotating, wasysym, url, multicol}
\usepackage[tight]{subfigure}
\usepackage[ruled,linesnumbered,vlined]{algorithm2e}
\usepackage{hyperref} 
\usepackage{breakurl}
\usepackage{color}
\usepackage[utf8]{inputenc}
\usepackage{tcolorbox}
\usepackage{caption}
\usepackage[english]{babel}
\usepackage{comment}
\usepackage[normalem]{ulem}
\usepackage{enumitem}
\usepackage{etoolbox}
\usepackage{graphicx,epstopdf}
\usepackage{tabularx}
\usepackage{float}

    \definecolor{zhao}{RGB}{205,133,0}

\usepackage[textsize=tiny,textwidth=0.6in]{todonotes}
\newcommand{\allnotes}[1]{}
\renewcommand{\allnotes}[1]{\textit{#1}}

\begin{document}

\date{}









\title{Quark: A High-Performance Secure Container Runtime for Serverless Computing}




\author{
{\rm Chenxingyu Zhao} \\
University of Washington
\and
{\rm Yulin Sun} \\
Quarksoft
\and
{\rm Ying Xiong} \\
CentaurusAI Inc
\and
{\rm Arvind Krishnamurthy} \\
University of Washington
}

\maketitle

\begin{abstract}
Secure container runtimes serve as the foundational layer for creating and running containers, which is the bedrock of emerging computing paradigms like microservices and serverless computing. Although existing secure container runtimes indeed enhance security via running containers over a guest kernel and a Virtual Machine Monitor (VMM or Hypervisor), they incur performance penalties in critical areas such as networking, container startup, and I/O system calls.

In our practice of operating microservices and serverless computing, we build a high-performance secure container runtime named Quark. 
Unlike existing solutions that rely on traditional VM technologies by importing Linux for the guest kernel and QEMU for the VMM, we take a different approach to building Quark from the ground up, paving the way for extreme customization to unlock high performance. 
Our development centers on co-designing a custom guest kernel and a VMM for secure containers.
To this end, we build a lightweight guest OS kernel named QKernel and a specialized VMM named QVisor. The QKernel-QVisor codesign allows us to deliver three key advancements: high-performance RDMA-based container networking, fast container startup mode, and efficient mechanisms for executing I/O syscalls. In our practice with real-world apps like Redis, Quark cuts down P95 latency by 79.3\% and increases throughput by 2.43x compared to Kata. Moreover, Quark container startup achieves 96.5\% lower latency than the cold-start mode while saving 81.3\% memory cost to the keep-warm mode. Quark is open-source with an industry-standard codebase in Rust.
\end{abstract}

\section{Introduction}
\vspace{-5pt}
Containerization is the de facto deployment manner for emerging cloud-computing paradigms such as microservices and serverless computing, offering significant benefits in aspects of portability, resource efficiency, and ease of scaling\cite{containerization, mahgoub2022orion,sartakov2022cap,qiu2020firm,van2022blackbox, Groundhog,liu2023doing,xu2023dirigo}. Under the hood of containerization, container runtime (or container engine) is the critical software component that creates and runs containers over host operating systems. 

\begin{figure}[htbp]
     \centering
 \includegraphics[width=0.87\linewidth]{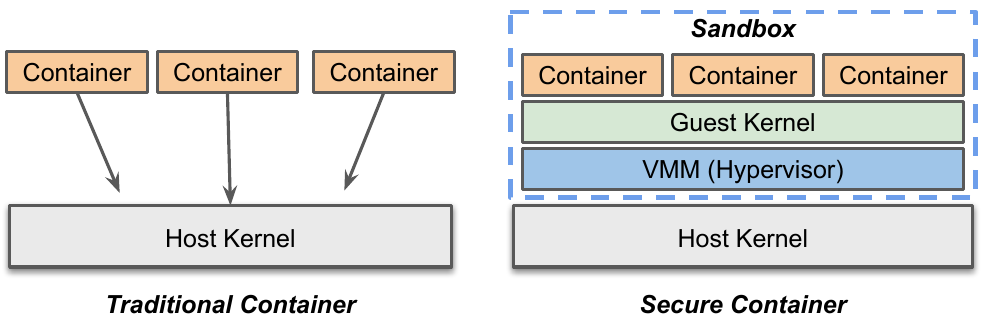}
  \vspace{-5pt}
 \caption{Traditional \textit{v.s.} Secure Container Runtimes.
 \vspace{-5pt}\label{fig::securemodel}}
 \end{figure}

In cloud providers' practice, secure container runtimes (also known as sandboxed runtimes) are widely used because of enhanced security, such as Google gVisor \cite{gVisor}, AWS Firecracker \cite{firecracker}, Alibaba runD \cite{RunD}, and Azure Sandboxing \cite{mspodsandbox}.  As Figure \ref{fig::securemodel} shows, the key to enhancing security in secure container runtimes lies in running containers on a userspace guest kernel of a lightweight Virtual Machine (VM), rather than directly on the host OS. The lightweight VM is launched by a Virtual Machine Monitor (VMM). Introducing the guest kernel and VMM adds an extra layer of isolation between containers and the host OS, reducing the risk of vulnerabilities (e.g., privilege escalation \cite{firecracker, cve2017, cve2019}). In common practice, the guest kernel is based on a full-ledge Linux kernel, and the VMM is based on the QEMU/KVM \cite{bellard2005qemu,kvm}, which is heavyweight (e.g., QEMU has over a million lines of code) and previously used for VM rather than containers.  

Although the involvement of guest kernel and VMM provides necessary security benefits, it introduces significant performance challenges. We indeed use existing secure runtimes like Kata \cite{kata} for customers' serverless platforms. However, we've identified the following key areas  where performance enhancements are desired to  improve the user experience:

$\bullet$ \textbf{High-performance Network:} Microservices and serverless computing rely heavily on networked systems.  On the one hand, complex monolithic services are decomposed into multiple containerized microservices connected via container networking.
On the other hand, client requests and service responses are also carried over the container networking. Unfortunately, existing secure container runtimes, which use kernel network stacks, often fail to meet high-performance networking requirements. The reason is that container communications might traverse both the guest kernel's and host OS's network stacks, both of which using Linux kernel have been shown to struggle with achieving high throughput and low latency \cite{zhang2021demikernel,zhang2022justitia,marty2019snap,guo2016rdma,dalton2018andromeda, zhuo2019slim, ma2022survey,su2022pipedevice}. As such, secure container runtimes should explore alternative, high-performance network stacks.

 $\bullet$ \textbf{Fast Startup:} In line with the microservices and serverless trend, containerized applications generally take the ephemeral execution, which frequently startup to serve requests and then quickly clean up after response. The container startup time takes a significant portion for response latency \cite{du2020catalyzer, yu2020characterizing, ustiugov2021benchmarking}, a critical factor for user experience.
 The cold-start approach results in prohibitively long latency (in the range of hundreds of milliseconds), while the keep-warm method leads to substantial memory overhead during idle time (practically running thousands of containers on a single machine \cite{firecracker,RunD}). It falls upon secure container runtimes to ensure that containers are booted both rapidly and efficiently in terms of memory cost. 

 $\bullet$ \textbf{Efficient Syscall:} Popular containerized applications for microservices, such as Redis \cite{redis} and Node.js \cite{nodejs}, are I/O-intensive and heavily rely on initiating I/O syscalls for disk or network access. Given the involvement of both the guest kernel and VMM, these system calls must traverse multiple layers. Typical secure container runtimes generally rely on the hypercall mechanism to trap into the VMM from the guest kernel. The conventional hypercall approach results in frequent context-switching between the guest kernel and the VMM, thereby leading to significant latency in I/O system calls. Therefore, secure container runtimes play a crucial role in enhancing the efficiency of syscall execution.

In our practice, we build \textbf{\textit{Quark}}, a high-performance secure container runtime that delivers significant improvements in the three critical areas previously discussed. The core principle behind Quark is the co-design of the guest kernel and the virtual machine monitor, which enables extreme customization to unlock high performance. 
To achieve such co-design, we build a customized guest kernel named \textbf{\textit{QKernel}} and a bespoke VMM named \textbf{\textit{QVisor}}. QKernel serves as the user-space guest kernel in a lightweight VM, encompassing subsystems like network stacks, memory management, process management, and the syscall virtualization layer, functionally similar to a typical OS kernel like Linux. QVisor functions as a virtual machine monitor (also known as a hypervisor), launching guest kernel and interacting with the host OS and physical devices on host machines. 

In the co-design of QKernel and QVisor, we highlight three critical advancements: 1) For container networking, we present \textbf{\textit{TCP Socket over RDMA}} (TSoR), which boosts network performance by transferring application TCP traffic over RDMA. Notably, TSoR requires no code modifications for containers that use standard POSIX socket APIs. 2) For container startup, we introduce the \textbf{\textit{Hibernation mode}}, which cuts down the startup latency while efficiently reducing the memory cost during idle times. Hibernation mode is enabled with a customized design for memory management and container snapshots. 3) For container syscalls, we design the \textbf{\textit{QCall}} mechanism, which speeds up the syscall by mitigating the overhead associated with context switching between the guest kernel and the VMM. Further details are in later sections: TSoR in $\S$\ref{sec::tsor}, Hibernation mode in $\S$\ref{sec::hibernate}, and QCall in $\S$\ref{sec::qcall}.

 Quark is developed with $\sim$135K lines of open-source, industry-standard code in Rust. Quark is compatible with Docker and Kubernetes. In $\S$\ref{sec::evaluation}, We evaluate Quark with end-to-end applications like Redis \cite{redis}, Node.js\cite{nodejs}, and Etcd \cite{etcd} and micro-benchmarks like iperf\cite{iperf}. Quark reduces the P95 latency of Redis by up to 79.3\% and boosts throughput by 2.43x compared to Kata \cite{kata}, a popular runtime.  For micro-benchmarks, Quark increases the iperf throughout by 2.46x and reduces the NPtcp latency by 86.6\%. Quark container startup achieves 96.5\% lower latency than the cold-start mode while saving 81.3\% memory cost to the keep-warm mode.
 More compelling results are in $\S$\ref{sec::evaluation}.

\section{Background and Motivation}
\label{sec::bg}
\subsection{Secure Container Runtime}
Secure container runtime enhances the security of traditional container runtimes (e.g., runC \cite{runc}) by adding an extra layer of isolation while running containers inside lightweight virtual machines (VMs), which is stronger than traditional containers’ processes-level isolation. Traditional container runtimes like runC \cite{runc} run containers sharing the same host kernel,  inherently leading to vulnerabilities for host OS such as container escaping attacks \cite{cve2019, cve2017}. To fortify the security boundary between the containerized processes and the host kernel, secure container runtimes employ a \textit{sandboxed} environment for containers. The combination of the guest kernel and the Virtual Machine Monitor (VMM) creates the sandboxed environment. One VMM process runs per guest kernel, while a group (pod) of containers could run over one guest kernel. Secure 
 runtime intercepts the system calls from containerized applications, thereby precluding direct interactions between the containers and the host kernel.

Secure container runtimes have seen widespread deployment \cite{kata,firecracker,gVisor,mspodsandbox,kuenzer2021unikraft}, with Kata~\cite{kata} serving as a widely-used example; Kata runs containers within a VM with a Linux-based guest kernel and utilizes QEMU as its default VMM. Although the sandboxed environment composed of the guest kernel and VMM effectively reduces security risks for the host OS, it introduces performance overhead for containers. The overhead is especially notable when the guest kernel is built on a heavyweight Linux kernel, and the VMM is built on QEMU — components originally designed for full-fledged VM virtualization as opposed to lightweight containerization. Next, we delve into the performance overhead.

\subsection{Performance Bottleneck}

\begin{table}[h]
    \footnotesize
    \centering
     \begin{tabular}{ c|c|c}
     \textbf{Network Solution} & \textbf{Latency} & \textbf{Throughput} \\
         \hline
         CNI based on Kernel TCP/IP& 64.09 us  & 17.1 Gbps \\
          \hline
         Quark based on RDMA& 8.97 us & 37.4 Gbps\\
        \end{tabular}
        \caption{Overhead of Kernel TCP/IP Stack}
        \label{tab:net}
\end{table}

 \textbf{Kernel Network Stack}: Currently, most of the secure container runtimes facilitate network connectivity by utilizing existing Container Network Interface (CNI) plugins such as Flannel \cite{Flannel}, Weave \cite{Weave}, and Cilium \cite{Cilium}. These CNI solutions build the data path based on the kernel TCP/IP stack. It is widely reported that the Linux kernel's TCP/IP stack incurs performance overhead, posing challenges for achieving high throughput and low latency \cite{zhang2021demikernel,zhang2022justitia,marty2019snap,guo2016rdma,dalton2018andromeda, zhuo2019slim, ma2022survey}. In the context of secure container runtimes, the overhead of kernel TCP/IP is particularly significant as network data typically traverses both the guest kernel's and the host kernel's network stacks, thereby amplifying inherent inefficiencies. As a result, high-performance networking solutions such as RDMA emerge as compelling alternatives for circumventing the bottlenecks associated with kernel-based network stacks. Some academic researchers have also begun harnessing RDMA to accelerate serverless computing \cite{wei2023no,wei2022krcore}. 
 As Table \ref{tab:net} shows, Quark can achieve lower latency and higher throughput while  replacing the kernel TCP/IP with the RDMA-based network solution (test setup in $\S$ \ref{sec::evaluation}).

\begin{table}[h]
    \footnotesize
    \centering
     \begin{tabular}{ c|c|c}
     \textbf{Test} & \textbf{Latency} & \textbf{Throughput} \\
         \hline
         Redis Ping w/ Hypercall & 11.0 $\mu s$/req & 90.1K RPS\\
          \hline
         Redis Ping w/ Qcall & 5.0 $\mu s$/req & 200K RPS \\
            \hline
        Hypercall Overhead (e.g., \textit{ sys\_read}) & 2.0 $\mu s$/req & - \\
        \end{tabular}
        \caption{Overhead of Hypercall}
        \label{tab:hypercall}
\end{table}

\textbf{Hypercall and Context Switch:} A hypercall serves as a trap mechanism that enables a guest VM to request privileged operations from the VMM.  The relationship between a hypercall and a VMM is similar to that between a syscall and an OS kernel. Given that secure container runtimes run containers on top of guest kernels and VMMs, they inherently introduce hypercall mechanisms into the system. However, hypercall requires costly context-switching between the guest kernel and VMM (e.g., switching registers and protection domains). Such context-switching overhead has been widely reported as a performance bottleneck in prior work \cite{prakash2022portkey, humphries2021case,zhou2023userspace,vmexit}. To measure the impact of this overhead on containerized applications, we conducted one micro-test with Redis, which is hypercall-intensive due to frequent I/O operations. Redis Ping invokes sys\_read, sys\_write, and epoll\_wait hypercalls. As Table \ref{tab:hypercall} shows, we can see that Redis Ping with non-optimized hypercall mechanisms requires about 11 us response latency. Specifically, hypercalls such as sys\_read incur as much as 2 us latency, significantly influencing both throughput and latency metrics. Here, we present a preview of results showing the efficacy of QCall—an optimized hypercall mechanism by minimizing context-switching overhead (further details and test setup are discussed in later sections), which effectively increases throughput and reduces latency.

\begin{table}[h]
    \footnotesize
    \centering
     \begin{tabular}{ c|c|c}
     \textbf{Startup Mode} & \textbf{Response Latency} & \textbf{Idle Memory Cost} \\
         \hline
         Cold Start & 563 ms  & - \\
          \hline
         Keep Warm & 1.4 ms & 40.82 MB\\
        \end{tabular}
        \caption{Overhead of Container Startup}
        \label{tab:startup}
\end{table}

\textbf{Container Startup: } The response latency for user requests to containerized services generally comprises three parts: container startup, application initialization, and user request processing. To measure the impact of container startup on response latency, we conduct a micro-test using float processing as a containerized service (further details and test setup are discussed in $\S$\ref{sec::evaluation}).  Table \ref{tab:startup} shows two modes of operating containers: Cold Start where containers are initialized only upon request arrival, and  Keep Warm where containers are pre-initialized and maintained in a ready state. Our results indicate that container startup can take several hundred milliseconds, significantly affecting overall response latency. However, while mitigating startup latency, the keep-warm approach incurs idle-time memory overhead, constraining the density of deploying containers on the same host.
In practical large-scale microservice deployments where a single machine may host thousands of containerized service instances \cite{firecracker,reap}, the idle memory overhead can become prohibitively huge. Therefore, there's a pressing need for a container startup mode that simultaneously minimizes both latency and memory overhead.

\section{Overview of Quark}

 \begin{figure}[htbp]
     \centering
 \includegraphics[width=0.97\linewidth]{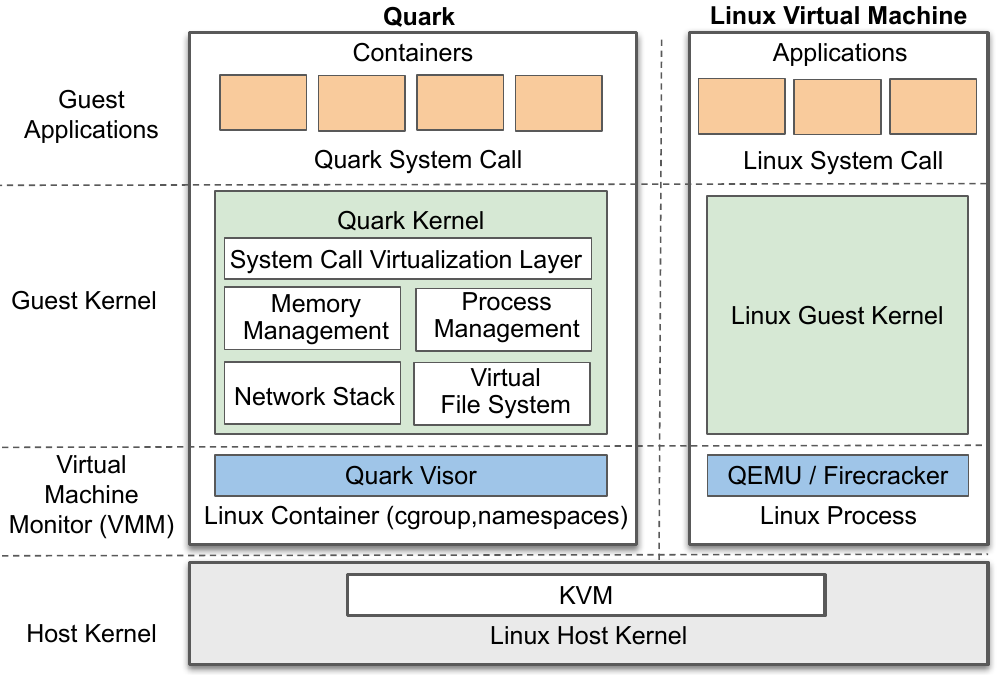}
 \caption{Quark secure container runtime \textit{v.s.} Linux VM.  \label{fig::quark-overview}}
 \vspace{-20pt}
 \end{figure}
 
\subsection{Principle: Co-design QKernel and QVisor}

In this paper, we present our experience about how we build a high-performance secure container runtime,  named \textbf{\textit{Quark}}.  
The key design principle for the Quark is to co-design the guest kernel and the virtual machine monitor (VMM). For this purpose, we design a specialized guest kernel, named \textbf{\textit{QKernel}}, along with a bespoke VMM, named \textbf{\textit{QVisor}}. As depicted in Figure \ref{fig::quark-overview}, QKernel is functionally equivalent to the Linux Kernel used in VM, and QVisor is equivalent to the QEMU.  Importantly, the ground-up co-design of QKernel and QVisor delivers tremendous opportunities to flexibly address the performance overhead commonly present in existing secure container runtimes. Next, we introduce the overview of QKernel and QVisor, two key components of the Quark secure container runtime. In later sections, we present high-performance mechanisms of Quark enabled by the QVisor-QKernel co-design.

\subsection{QVisor: Hypervisor/VMM}

QVisor (short for Quark Visor) is a VMM (also known as a hypervisor) that creates and manages lightweight VMs to run containers. Similar to QEMU, a typical VMM, QVisor is to allocate and isolate host resources such as CPU, memory, and network interface. However, QVisor is more lightweight than QEMU, as it discards many heavyweight but non-essential features for containers, such as device emulation. More importantly, QVisor is co-designed with QKernel, facilitating targeted optimizations in functionalities like hypercalls, network stacks, and container startup, thereby surpassing the capabilities of typical VMMs like QEMU while serving sandboxed containers. 

QVisor is implemented as a Linux Container rather than as a simple process running atop the host kernel to conveniently utilize kernel features: cgroup and namespace. Cgroup (control group) allows the QVisor to enforce fine-grained resource limitations on host resources like CPU, memory, and networking. Meanwhile, namespaces provide QVisor with the capability to instantiate isolated views of global host resources such as hostnames, network namespaces, and file systems.
The combined use of cgroups and namespaces facilitates enhanced isolation and resource management capabilities for running multiple sandboxed containers on the same host.

\subsection{QKernel: Guest Kernel}

QKernel (short for Quark Kernel) functions as the guest OS kernel within a lightweight VM. As depicted in Figure \ref{fig::quark-overview}, QKernel encompasses multiple subsystems similar to those in the Linux Kernel—these include syscall virtualization layers, memory management, process management, a virtual file system, and a custom network stack: 1) The System Call Virtualization Layer of Quark supports POSIX-compliant syscalls for containers. This ensures compatibility with a broad range of container images without requiring any modifications to user code. Notably, standard TCP POSIX socket APIs such as \textit{connect()}, \textit{accept()}, and \textit{send()} are fully supported. 2) Under the hood of socket APIs, QKernel features its high-performance network stack, named TCP Socket over RDMA, which is fundamentally different from Linux Kernel's TCP/IP stack. Further details are covered in $\S$\ref{sec::tsor}. 3) The Process Management subsystem sets up vCPUs by leveraging the KVM facilities in the host kernel. It also handles hypercalls interacting with QVisor. Here, we introduce a more efficient hypercall mechanism, named QCall, made possible by the co-design of QKernel and QVisor. Detailed information is provided in $\S$\ref{sec::qcall}. 4) Memory Management involves allocating the memory provisioned by QVisor and handling the address translation between guest containers and host memory. This subsystem plays a pivotal role in optimizing container startup times, as detailed in $\S$\ref{sec::hibernate}. 5) The Virtual File System (VFS) is to provide the filesystem interface for containers by handling various types of IO operations. QKernel fully supports access to virtual files such as \textit{/dev} and \textit{/proc}. For physical files, QKernel supports container access to a host directory tree through file system passthrough.

\section{Container Network: TSoR}
\label{sec::tsor}
\subsection{Rationale of Co-design}

The co-design of QKernel and QVisor offers an opportunity to create a highly efficient network stack that spans from the NIC's device layer up to the TCP socket layer. Such co-design is advantageous in two key respects. On the one hand, QKernel handles system call virtualization, including support for POSIX socket APIs, laying the groundwork for transparently boosting network performance without requiring any code changes in the applications. Currently, most containerized applications use standard POSIX sockets for communications, regardless of using synchronous manners such as HTTP/gRPC\cite{grpc} or asynchronous manners such as AMQP~\cite{amqparch}. On the other hand, QVisor can directly access the physical NIC on host machines and even utilize kernel-bypass techniques such as RDMA. Leveraging these capabilities, we introduce an efficient network solution called \textit{TCP Socket over RDMA}. This solution transparently accelerates applications that use POSIX socket interfaces by taking advantage of RDMA's low-latency, high-bandwidth, and low CPU utilization benefits.

\subsection{TCP Socket over RDMA}

\begin{figure}[ht!]
     \centering
\includegraphics[width=0.8\linewidth]{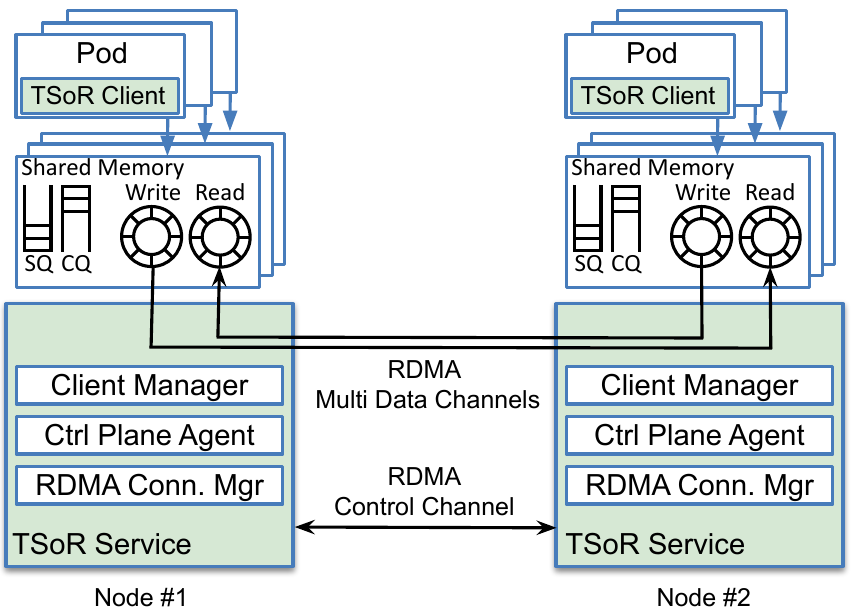}
 \caption{TSoR Architecture}
 \label{fig::rdma-service}
 \end{figure}

TSoR (short for TCP Socket over RDMA) is the networking solution of Quark container runtime. Applying the model of the \textit{network-stack-as-a-service} \cite{niu2017network,niu2021netkernel,bastion}, TSoR comprises two essential service/client components shown in Figure \ref{fig::rdma-service}: 1) TSoR Service is the core component that provides the network connectivity for the container pods. A pod is a group of one or more containers running over the same QKernel with shared resources such as network namespace and IP address. TSoR Service manages RDMA connections and executes data transmission to other machines via RDMA NIC. TSoR Service also talks to the network orchestration control plane, e.g., Kubernetes API Server. 2) Each pod has a TSoR client. After QKernel intercepts containerized applications' POSIX socket API calls, the TSoR client will take over and talk to TSoR Service to set up the connection with peer nodes and transmit data over RDMA. In the following sections, we describe the TSoR client, TSoR service, and network operations they facilitate.

\subsection{TSoR Client}
\label{sec::tsor-client}

The primary role of the TSoR Client is to intercept socket API calls from containerized applications. QKernel implements all system calls required by POSIX socket APIs,  including two types: 1) Control primitives such as \textit{connect}, \textit{listen}, \textit{accept}, and \textit{close} are to set up/tear down the
TCP connections; 2) Data primitives such as \textit{read}, \textit{write}, \textit{send}, and \textit{recv} are used to process data transmission. In $\S$ \ref{sec::tsoroperation}, we will describe how POSIX socket API calls are processed from the end-to-end workflow. 

TSoR Client takes over socket system calls and then communicates with TSoR Service using Shared Memory. Each client creates one shared memory region, which consists of two key data structures: 1) \textit{Message Queue Pair}: The message queue pair consists of two shared memory queues: Submission Queue (SQ) and Completion Queue (CQ) \footnote{Here, SQ and CQ are used for communication between TSoR Client and TSoR Service, which are different from Queue Pairs of RDMA connection.}. TSoR Clients send request messages to TSoR Service through SQ and receive response messages from TSoR Service through CQ.
2) \textit{Shared Buffers}: Shared Buffers consist of read/write data buffers to store data for application send()/receive(). Shared buffers play a similar role as the socket buffers of standard TCP/IP stack. Note that shared buffers between the TSoR Client and TSoR Service also serve as the registered Memory Region (MR) for RDMA.

\subsection{TSoR Service}
\label{sec::tsor-serivce}

In this section, we introduce the submodules of the TSoR Service, including the Client Manager, Control Plane Agent, and RDMA connection manager. Among them, the RDMA connection manger is the core submodule. 

\subsubsection{Client Manager}

The Client Manager is responsible for managing TSoR clients and facilitates communication with them through shared memory regions. The shared memory region maintains metadata and data structures associated with each TSoR client, which is described in the TSoR Client $\S$\ref{sec::tsor-client}.


\subsubsection{Control Plane Agent} TSoR Service needs to get connection-related metadata from the orchestration system control plane, e.g., Kubernetes API Server in the Kubernetes cluster. The connection-related metadata includes active cluster node list, Pod list, and cluster connection permission control policy. Based on the metadata, TSoR Service determines whether it is permitted to and how it sets up virtual TCP connections (which we refer to as an RDMA Data Channel) to map to real TCP connections in TSoR Clients.

\subsubsection{RDMA Connection Manager} 
\label{sec::rdma-cm}
RDMA Connection Manager handles RDMA Queue Pair (QP) connections, which use Reliable Connection (RC) transport modes\cite{rdmaprogamming}. RDMA Connection Manager is responsible for creating and cleaning up QP connections. 
QP connection is used to transfer TCP traffic over RDMA. There are two main challenges for RDMA Connection Manager design:

\textbf{Challenge \#1: RDMA connection scalability}. In a typical cluster environment, micro-services within containers could set up a large number of concurrent TCP connections. However, the existing RDMA network falls short on the scalability issue, which is widely reported in \cite{kalia2016design, kong2022collie, wang2019vsocket, zhang2022justitia}. When the number of QP connections is large, the aggregated performance degrades dramatically. The root cause is the contention on the RDMA NIC's internal hardware resource, which is beyond the control of the container network. Given the high concurrency of scenarios using TCP connections, it is not practicable to build a one-to-one mapping between the TCP connection and the RDMA QP connection. 

\textbf{Challenge \#2: RDMA connection setup latency}: RDMA connection setup consists of creating resources (e.g., QP), exchanging metadata information, and changing state. Usually, RDMA uses a TCP connection as a communication channel to exchange QP metadata. We measured the latency of creating a QP Connection based on TCP and found the latency is up to several milliseconds. For scenarios with frequently establishing short-lived TCP socket connections, the time cost is significant if every short-lived TCP connection requires creating a separate RDMA connection.

We provide two solutions to tackle the aforementioned challenges together:

\textbf{Solution \#1: Multiplex Node-level RDMA connection}.
To solve the RDMA connection scalability issue, we multiplex the single long-lived RDMA connection for all TCP connections between the same pair of nodes.  Thus, the number of RDMA connections is determined by the number of cluster nodes, which is orders of magnitude less than the number of TCP connections. RDMA connection multiplexing enables TSoR to support a large number of concurrent TCP connections.
 As Figure \ref{fig::rdma-service} shows, TSoR introduces a concept of RDMA Channel, which is mapped to a TCP socket connection.  Each end of the RDMA channel has two data ring buffers: a read buffer and a write buffer.   
  The write buffer on one end connects to the read buffer on the other end. Besides RDMA Data channels for data transfer, each pair of nodes maintains one RDMA Control Channel to exchange control messages. For example, nodes use control messages to tell remote peers about the available space of read buffer for enforcing rate control.

\textbf{Solution \#2: Pre-established RDMA Connection}. Due to the long latency of RDMA connection setup, TSoR pre-establish RDMA connections rather than doing that when TCP connections are requested by user applications. When a node joins the cluster, the TSoR Service on the node will start up and establish RDMA Connections to all its peer nodes in the cluster. At the time applications initiate TCP connections, the RDMA data channel can directly use the pre-established RDMA connection. By using the pre-established RDMA connection, TCP connection establishment does not need to pay the time cost for RDMA QP setup. Also, the low latency and reliable data path of RDMA can speed up the handshake process of TCP. In $\S$\ref{sec::handshake}, we will describe the detailed process of TCP connection establishment.

\subsection{TSoR Operation}
\label{sec::tsoroperation}

\begin{figure}[tbp]
     \centering
 \includegraphics[width=\linewidth]{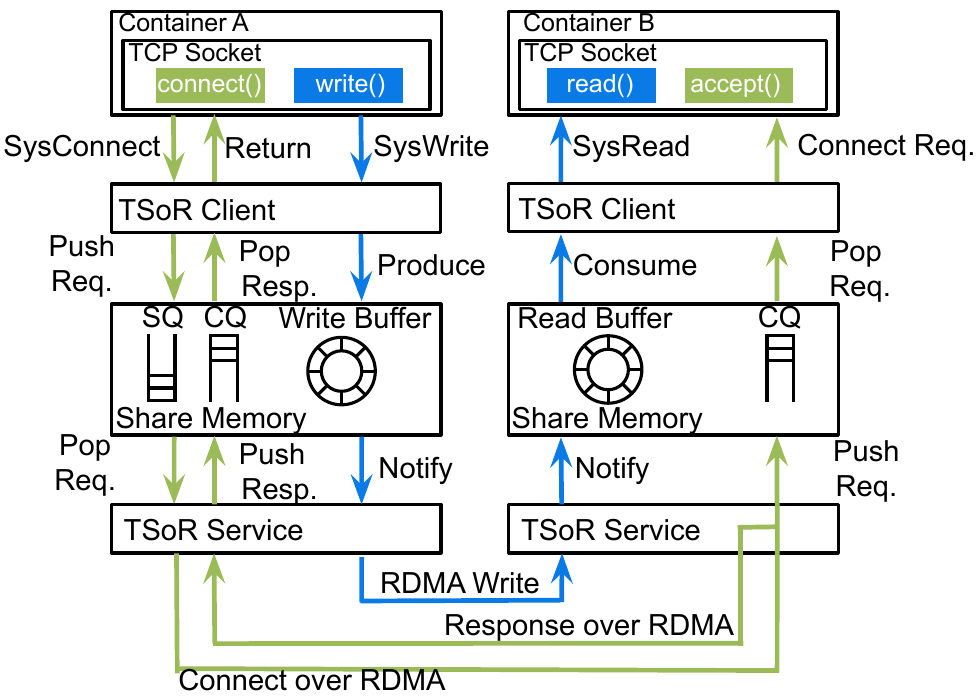}
 \caption{TCP data transmission and connection setup.}
 \label{fig::data-transmission}
 \end{figure}

\subsubsection{Data Trasmission}
Figure \ref{fig::data-transmission} shows the workflow of how two containers on different cluster nodes use TSoR to write()/read() data, which has the following main steps: 

$\bullet$ \textbf{Step \#1}: Containerized application calls $write$ which invokes the system call $SysWrite$. The TSoR client intercepts the system call and interacts with the TSoR service. The TSoR Client plays as the producer for the write buffer by copying data from the application into the write buffer.

$\bullet$ \textbf{Step \#2}: TSoR clients enqueue one Write Request to the submission queue shared between the TSoR client and TSoR Service. Write Request is a signal to trigger the process of TSoR service. 

$\bullet$ \textbf{Step \#3}: TSoR service dequeues the Write Request and transmits data to the peer node over RDMA connection if the remote read buffer still has space. TSoR Service plays as a consumer for the local write buffer by transferring data to the remote peer's read buffer using RDMA IB verb.

$\bullet$ \textbf{Step \#4}: When TSoR Service is notified upon the arrival of data through RDMA completion event, it enqueues one request into the CQ between the TSoR Service and TSOR client to indicate the read buffer has new arrival data. 

$\bullet$ \textbf{Step \#5}: When the TSoR client is notified that the read buffer has data arrival, the TSoR client will consume data by copying data from the read buffer to the application buffer. The application finally receives the data from the remote peer.

We make several optimizations for the above data transmission workflow:

$\bullet$ \textbf{Optimization \#1 Pipeling}: To enhance data transfer efficiency, we implement pipelining in the Producer-Consumer workflow. On the sender side, the TSoR client fills the write buffer as the producer, while the TSoR service reads this buffer and transfers data to the receiver. Conversely, on the receiver side, the TSoR service acts as the producer for the read buffer, and the TSoR client serves as the consumer.
On both the sender and receiver sides, the producer and consumer can work pipelining while transferring a stream of data packets. This pipelining increases data transfer throughput for maximizing the utility of high-throughput RDMA connections.

$\bullet$ \textbf{Optimization \#2 Signal Coalescing}: To minimize transmission overhead, we introduce \textit{Signal Coalescing}. This approach allows TSoR clients to skip enqueuing the Write Request to notify the TSoR service to process when they find the write buffer already contains data. Correspondingly, the TSoR Service checks the ring buffer after each RDMA operation and continues to send the remaining data without requiring a new Write Request.  Via the \textit{Signal Coalescing}, TSoR client and service can save the cost of manipulating the SQ for most cases. SQ is only used to initiate the data transfer.         
    
$\bullet$ \textbf{Optimization \#3 Idle Sleep and notification bitmap}: 
 We enable the mechanism of idle sleep for TSoR Service. In Step-3, instead of continuously polling the submission queue, the service uses a hybrid mode that combines busy polling with event notification. In idle periods, the TSoR Service sleeps until awakened by an event notification from a client. After waking up, it reverts to busy polling for a brief period before returning to sleep mode. To manage multiple client requests efficiently, we employ a 2-layer bitmap, enabling rapid identification of the requesting TSoR client, thus further enhancing performance.

$\bullet$ \textbf{Optimization \#4 Lazy notification for read buffer available space}: In Step-3, the TSoR service needs to check whether the remote read buffer has available space before sending data over RDMA.  If the buffer is full, the service temporarily stops data transfer. When the application consumes data in the read buffer, the available space will increase, and the TSoR service will notify the other side with the new space using the RDMA control channel.  However, to avoid performance degradation due to excessive notifications, we set a threshold: notifications are only sent when more than half of the total ring buffer space becomes available.

\subsubsection{TCP Connection Establishment}
\label{sec::handshake}

In this section, we explain how TSoR clients establish TCP connections. Unlike the typical three-way handshake in standard TCP, TSoR uses a more efficient two-way handshake. This is made possible by leveraging pre-established RDMA connections between nodes. With the lower network latency of RDMA, TSoR can quickly establish TCP connections, thus reducing connection setup time.

Figure \ref{fig::data-transmission} shows the process of establishing a TCP connection with a two-way handshake as follows:    

$\bullet$ \textbf{Step \#1:} $Container~A$ initiates a TCP connect request by calling $connect$ which invokes system call $SysConnect$. The TSoR client intercepts $SysConnect$, obtains the destination IP address/port, and then interacts with the TSoR Service via enqueuing one TCP connect request into SQ. 

$\bullet$ \textbf{Step \#2:} TSoR Service running on $Container~A$'s host pops out the requests from SQ and extracts the destination container IP and port. TSoR Service looks up the peer cluster node hosting $Container~B$, creates a local RDMA Data Channel, and then sends a connection request using RDMA Control Channel over RDMA connection.  

$\bullet$ \textbf{Step \#3:} $Container~B$ waits on connection request via $accept$ socket API call. After the TSoR Service running on $Container~B$'s host receives the connection requests, two tasks are executed to accept a connection request: Firstly, the TSoR service enqueues a Connect Request into the CQ to inform the $Container~B$ to establish a new connection.  Secondly, the TSoR Service creates an RDMA Data channel and sends the response with read ring buffer information associated with the newly created RDMA Channel back to the peer node. 

$\bullet$ \textbf{Step \#4:} $Container~A$'s TSoR Service receives the response and then enqueue a TCP accept request into CQ to notify $Container~A$ that one connection is established and is ready for data transmission.

$\bullet$ \textbf{Step \#5:} TSoR Client in $Container~A$ pops out the response from CQ and then finishes the connection setup phase. The application is notified to be ready to send data.

\begin{figure*}[htbp]
	\centering
    \subfigure[Hibernation Mode]{
		\begin{minipage}[t]{0.3\linewidth}{
				\includegraphics[width=\linewidth]{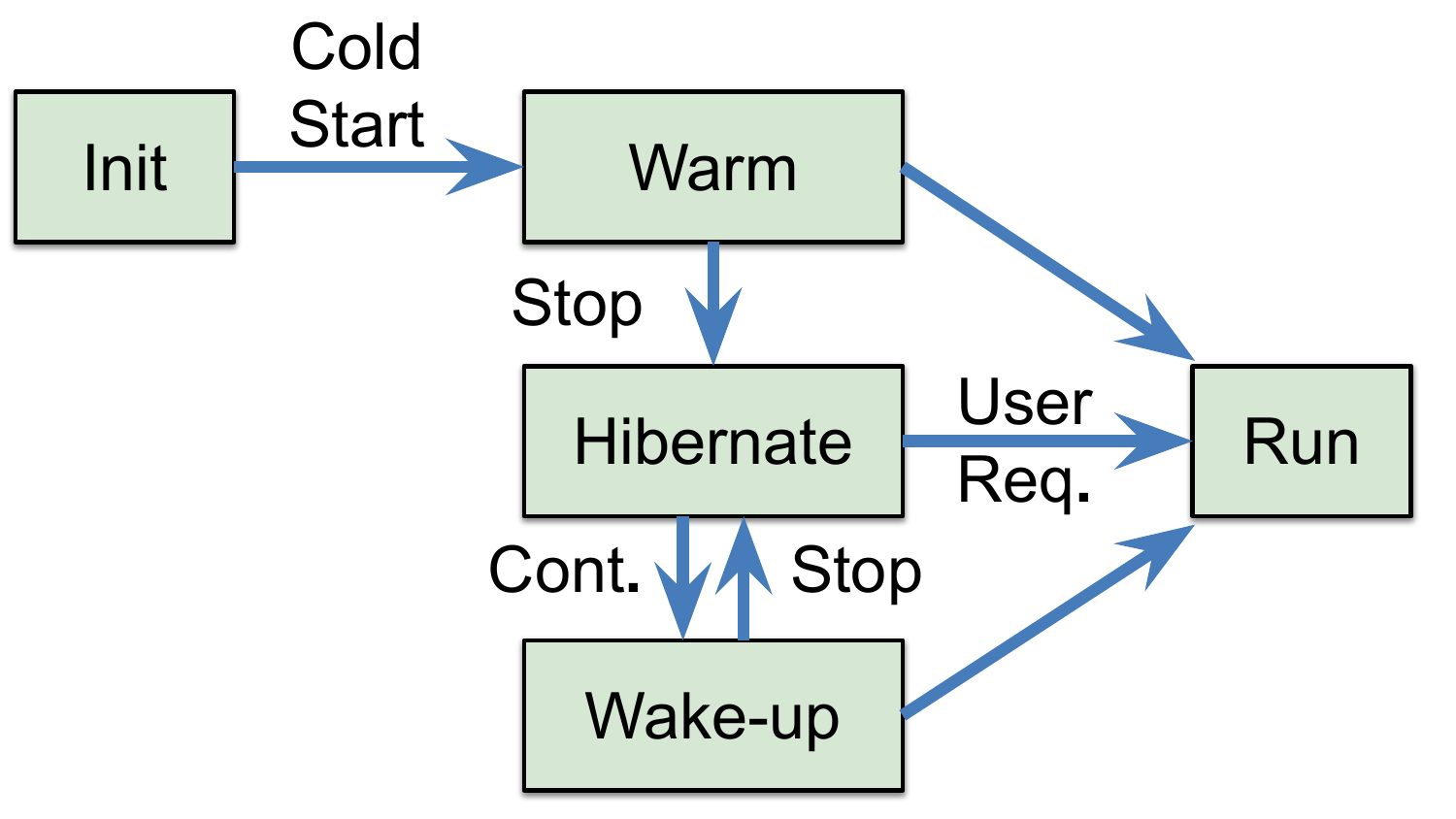}}
            \label{fig::hibernate}
    	\end{minipage}}
    \subfigure[Bitmap Page Allocator]{
		\begin{minipage}[t]{0.3\linewidth}{
				\includegraphics[width=\linewidth]{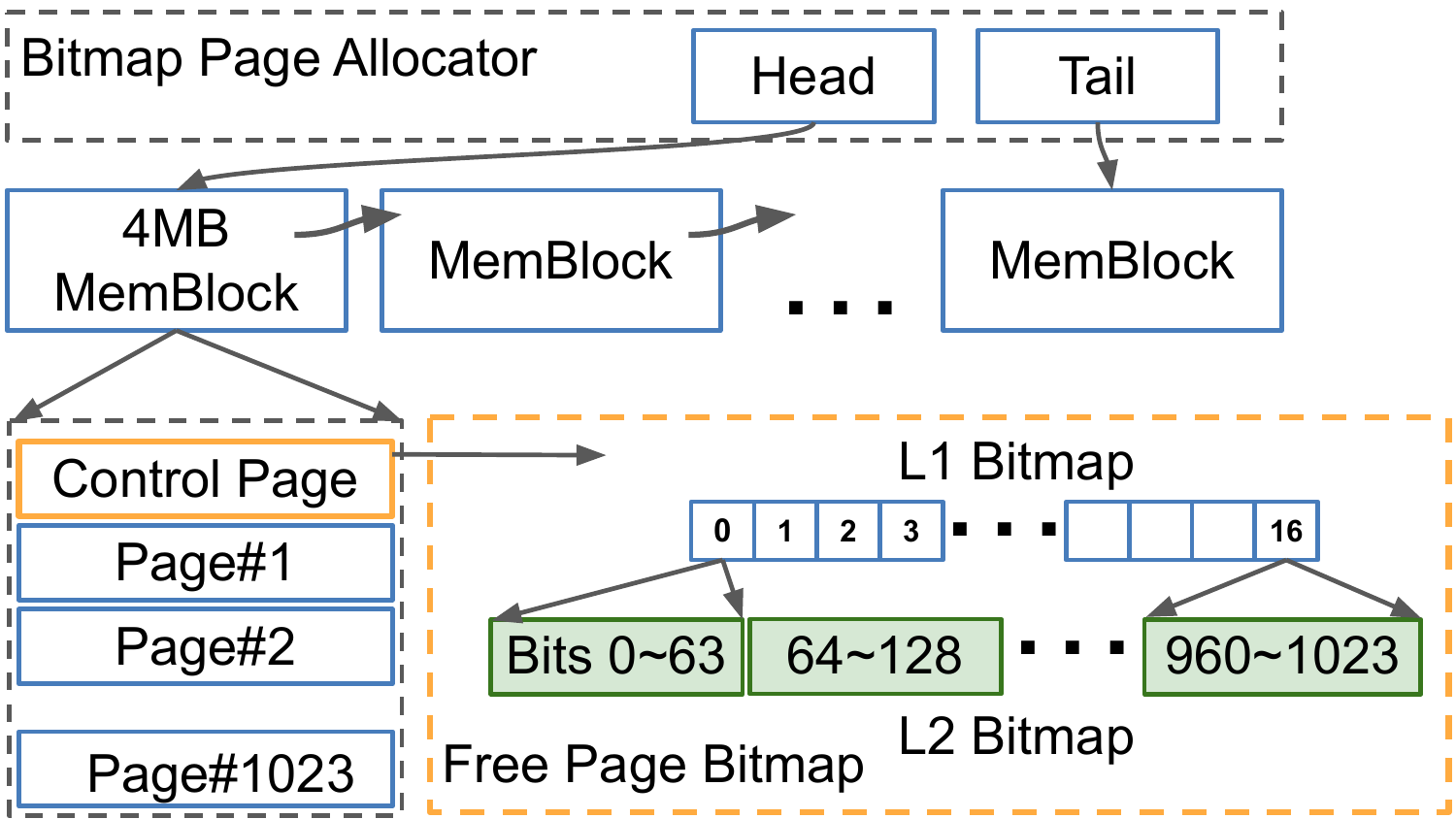}}
            \label{fig::pageallocator}
    	\end{minipage}}
    \subfigure[Memory Swapping]{
    \begin{minipage}[t]{0.3\linewidth}{
            \includegraphics[width=\linewidth]{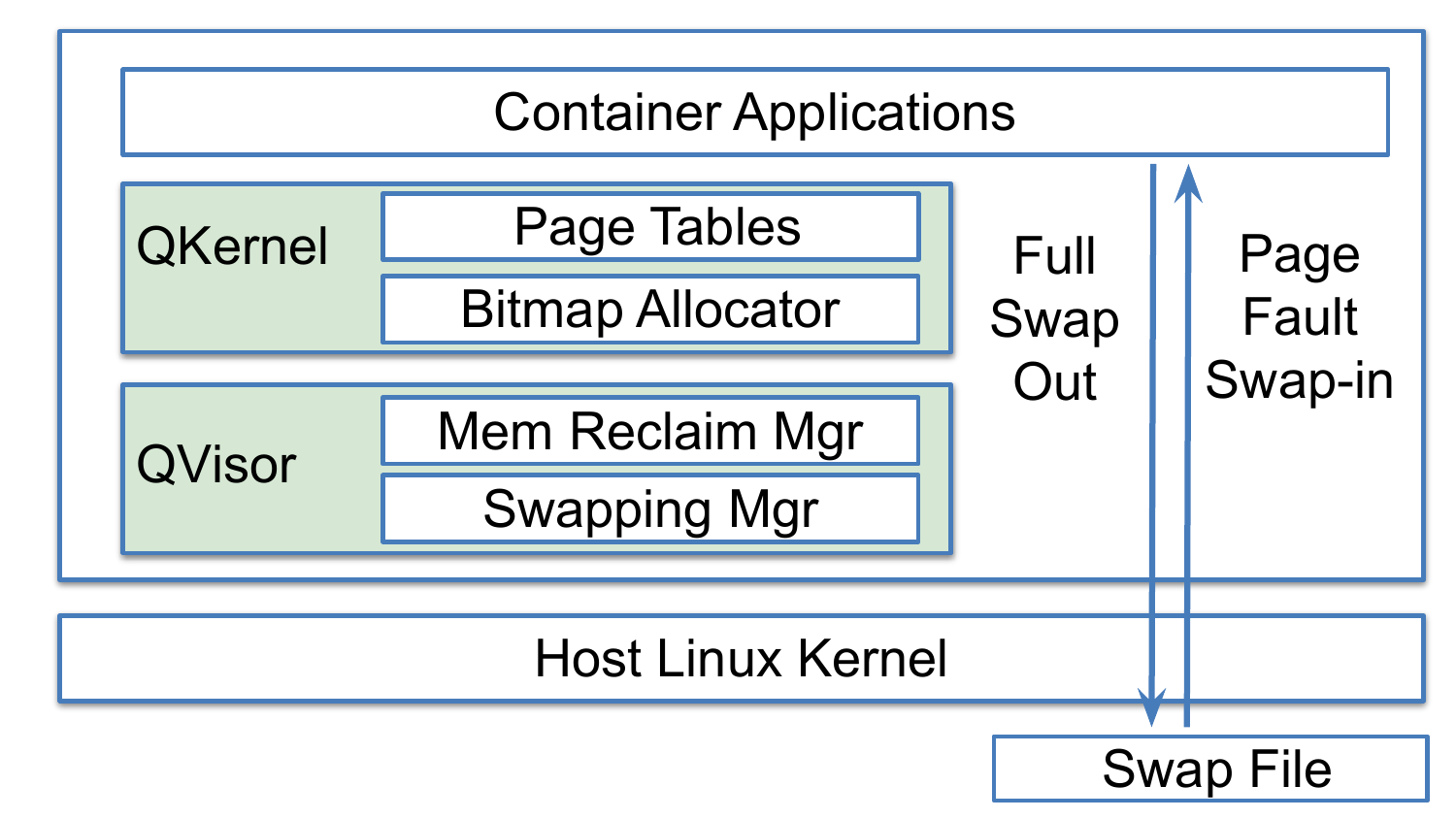}}
        \label{fig::swapping}
    \end{minipage}}
\caption{Quark enables Hibernate Mode via customizing the memory management and swapping. }
\label{fig::startup}
\end{figure*}

\section{Container Startup: Hibernation Mode }
\label{sec::hibernate}
\subsection{Rationale of Co-design}

The co-design of QKernel and QVisor introduces a highly efficient container startup mode called \textit{Hibernation Mode}. Traditional container startup methods suffer from trade-offs: the cold-start approach incurs high latency, whereas the keep-warm method results in excessive idle-time memory consumption. The Hibernation mode aims to reconcile these drawbacks by preserving essential runtime objects—such as container processes, cgroups, and file system mappings—while deflating the memory footprint of idle containerized applications.
Maintaining essential runtime objects in an active state minimizes startup latency while deflating memory during idle periods helps save memory usage.
This unique mode is made possible through two critical mechanisms—memory reclaiming and memory swapping—both of which necessitate tight coordination between QKernel and QVisor. For memory reclaiming, QKernel handles the memory allocation, while QVisor hosts the manager for memory reclaiming.  For memory swapping, QKernel requires customized page tables, while QVisor employs a dedicated swapping manager. Next, we delve into how such codesign delivers the Hibernate mode. 

\subsection{Hibernation Mode}

The Quark system introduces a hibernation mode that balances rapid container startup with efficient memory utilization. As Figure \ref{fig::hibernate} illustrates, Quark containers can exist in one of five operational states:
1) Init: Container runtimes, including QKernel and QVisor, are not yet launched, and containerized applications remain uninitialized.
2) Warm: The container runtime process is active, and the containerized applications have been initialized. Containers in this state are ready to handle incoming user requests.
3) Running: User requests arrive, and containers are busy processing the requests.
4) Hibernation: Containers in hibernation mode are \textit{deflated} warm containers.  Deflation here implies that the application processes are paused, and the associated memory is released back to the host kernel.
5) Wake-up: Containers in hibernation can be \textit{inflated} by restoring application memory and resuming application processes. Once awakened, these containers are ready to process incoming requests. Unlike containers in the warm state, those in the wake-up state employ on-demand memory swapping (more details in $\S$\ref{sec::memoryswap}).

To enable the transition from warm containers to hibernation mode, the following key steps are involved:
1) Pause application processes of containers and block the threads of container runtime. 
2) Reclaim application memory pages and release them back to the host OS, which is described in $\S$\ref{sec::reclaim}.
3) Create snapshots of the container's committed memory pages by swapping them out to disk storage, which is described in $\S$\ref{sec::memoryswap}.

\subsection{Memory Management and Reclamation}
\label{sec::reclaim}

In hibernation mode, Quark reclaims unused application memory pages and returns them to the host machine, thereby reducing memory overhead while containers are in hibernation. To accomplish this, Quark employs the system call \textit{madvise()} with the advice parameter set to \textit{MADV\_DONTNEED}. This informs the host kernel that the application does not anticipate using these specific memory pages in the near future, enabling the host machine to reclaim these pages to other processes. After executing the \textit{madvise()} call, the affected pages become zero-fill-on-demand pages, which are filled with zeros before being given to a process.

To enable the above procedure of reclaiming the memory, the page allocator within the QKernel needs a customized design. Typical memory allocators used by Linux kernels, such as Slab Allocator \cite{bonwick1994slab} and Buddy Allocator \cite{knowlton1965fast}, maintain the free memory blocks in a free list, which is a linear linked list, and its \textit{next} pointer is kept in the free memory blocks. The free list works well for the host OS kernel.  Unfortunately, such a free list design is not suitable for Quark to reclaim the free memory page blocks. Because when we use \textit{madvise()} to reclaim the free memory page blocks, as pages of the memory block are zero-filled, the \textit{next} point is cleared, so the free list data structure is broken. Thus, we propose Bitmap Page Allocator to address this particular issue.

As shown in Figure \ref{fig::pageallocator}, the Bitmap Page Allocator manages fixed-size memory blocks, each consisting of 1024 memory pages that are 4KB in size. The first page in each block serves as the control page, maintaining a two-layer hierarchical bitmap to indicate whether each page is free.  The L1 bitmap, with 16 bits, narrows the search range by indexing the L2 bitmap, which has 1024 bits. Each bit of the L2 bitmap indicates the allocation status for one page (except for the control page).
Each bit of the L1 bitmap corresponds to 64 bits in the L2 bitmap, marking the presence of at least one free page within the indexed 64 pages. To locate a free page, the allocator first identifies a non-zero bit in the L1 bitmap and then scans the corresponding 64 bits in the L2 to locate the exact free page.

\subsection{Snapshot by Memory Swapping}
\label{sec::memoryswap}

Before transitioning the container into hibernation mode, Quark needs to \textit{snapshot} application data in memory by swapping the data in memory to the persistent storage, such as local disks, in preparation for future restoration. As shown in Figure \ref{fig::swapping}, the Swapping Manager in QVisor is responsible for carrying out the swap-out procedures. 
Conversely, when a container is woken up, the Swapping Manager executes the corresponding swap-in operations. 
Next, we detail the swap-out and swap-in mechanisms, respectively.

The swap-out process unfolds as follows: 1) QKernel pauses application threads, allowing for the memory pages used by the application to be swapped out. 2) The Swapping Manager scans the guest application's page tables and marks each anonymous page's entry as \textit{Not-Present}, which will cause a page fault upon future access.  The page fault serves as a trigger for a subsequent swap-in operation, the details of which will be discussed later. 3) The memory pages are then written to a swap file stored on disk. 4) Finally, the swapped-out pages are returned to the host OS through a memory reclamation procedure described before.

The swap-in process unfolds as follows: 1) When a hibernated container is reactivated, accessing a swapped-out memory page triggers a page fault, starting the swap-in process. 2) The page fault handler, executing on the vCPU, traps from QKernel to QVisor to read the required memory page from the swap file. 3) The page table entry is then marked as \textit{Present}, preventing further page faults for that particular page. 
Note that the swap-in process is conducted in an on-demand manner rather than loading all data from disk to memory all at once when the container is woken up.

\section{Container I/O Syscall: QCall}
\label{sec::qcall}

\subsection{Rationale of Co-design}

The co-design of QKernel and QVisor unlocks an opportunity to implement a hypercall mechanism from the ground up, addressing the traditional overhead associated with context switches between the guest kernel and the VMM. To this end, we introduce \textit{QCall}, which allows the QKernel to request privileged operations from QVisor.  QCall retains the functionalities of traditional hypercalls while minimizing the overhead of context-switching.

\begin{figure}[htbp]
     \centering
 \includegraphics[width=0.9\linewidth]{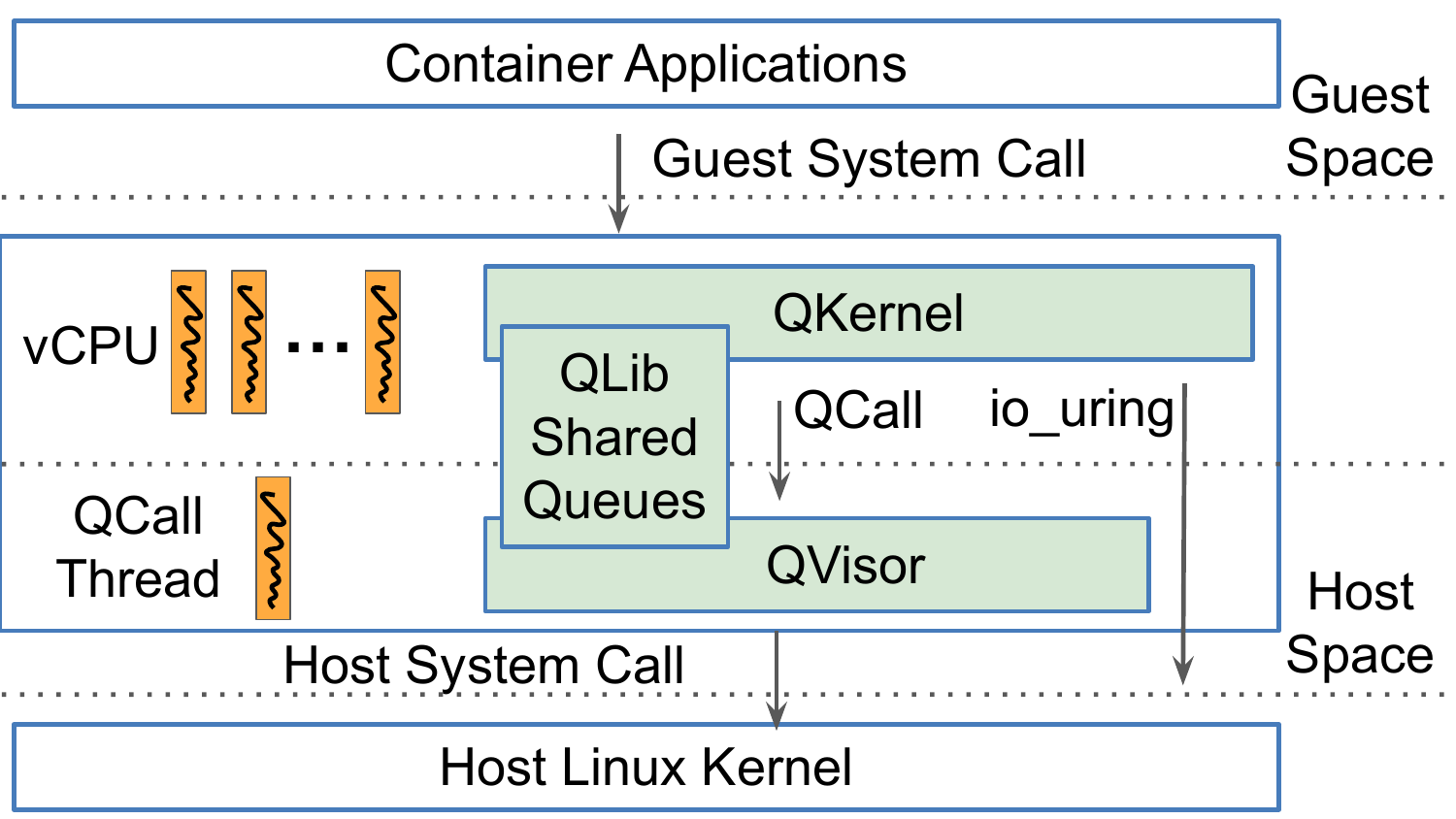}
 \caption{QCall mechanism}
 \label{fig::qcall}
\end{figure}

\subsection{QCall}

QCall (short for Quark Call) serves as a mechanism that allows the QKernel to execute privileged operations via QVisor. Diverging from traditional hypercall mechanisms, which typically employ a synchronous workflow by trapping into the VMM, QCall opts for an asynchronous approach. As Figure \ref{fig::qcall} shows, QCall is facilitated through two core components: 
1) \textit{QLib} is a library that establishes shared queues between QKernel and QVisor via memory sharing. These queues accept jobs submitted by the vCPU threads running in QKernel.
2) \textit{QCall Handling Thread}, a dedicated thread waiting in QVisor, is responsible for dequeuing jobs from QLib and executing the corresponding requests. 

The workflow for virtual threads requesting privileged operations unfolds as follows: Virtual threads in QKernel submit a job into QLib, subsequently entering a blocked state as they await job completion. Importantly, the QKernel then swaps out these blocked threads from the vCPU, allowing the vCPU to execute other runnable threads—ensuring the vCPU itself is never blocked.  The QCall handling thread will dequeue and execute the job from QLib.  Upon completion, QVisor notifies QKernel, transitioning the waiting threads to a \textit{ready-to-execute} state. The vCPU will then execute these threads.  Note that QLib doesn't take the design of a completion queue to avoid the cost of virtual threads polling for a completion signal.
Once the job is completed, the thread is transitioned to a ready-to-execute state and is executed by the vCPU as scheduling allows.

One of the advantages of QCall is the elimination of the need for vCPU to trap into QVisor's context from the QKernel's context. Because virtual threads invoke privilege operations via submit jobs rather than trapping into the QVisor. This effectively removes the context-switching overhead associated with traditional hypercalls. In traditional hypercalls, a trap into the VMM is required, followed by a context switch back to the guest kernel after execution is complete. Furthermore, The async workflows allow the vCPU to execute other threads without being blocked or engaged in busy polling.

\subsection{Host IO Operations via io\_uring}
For host I/O operations like file read/write, QKernel leverages io\_uring \cite{iouring} to directly interact with the host kernel, bypassing the QCall mechanism to further accelerate I/O performance. io\_uring is an advanced asynchronous I/O interface in the Linux kernel that is designed to reduce overhead by enabling batched and asynchronous I/O requests. It employs a set of lock-free ring buffers that are shared between user and kernel spaces to handle the submission and completion of I/O tasks. This efficiency enables Quark to swiftly execute I/O operations. Importantly, while Quark uses io\_uring for data access operations to enhance performance, it maintains control operations such as file opening through QVisor to enhance security. For example, opening a file generally involves setting permission controls, such as readability and writability. Also, control operations are generally less amenable to batching compared with data operations, so performance improvements from io\_uring are limited for control operations.

\begin{figure*}[ht!]
	\centering
    \subfigure[Single Connection]{
		\begin{minipage}[t]{0.3\linewidth}{
				\includegraphics[width=\linewidth]{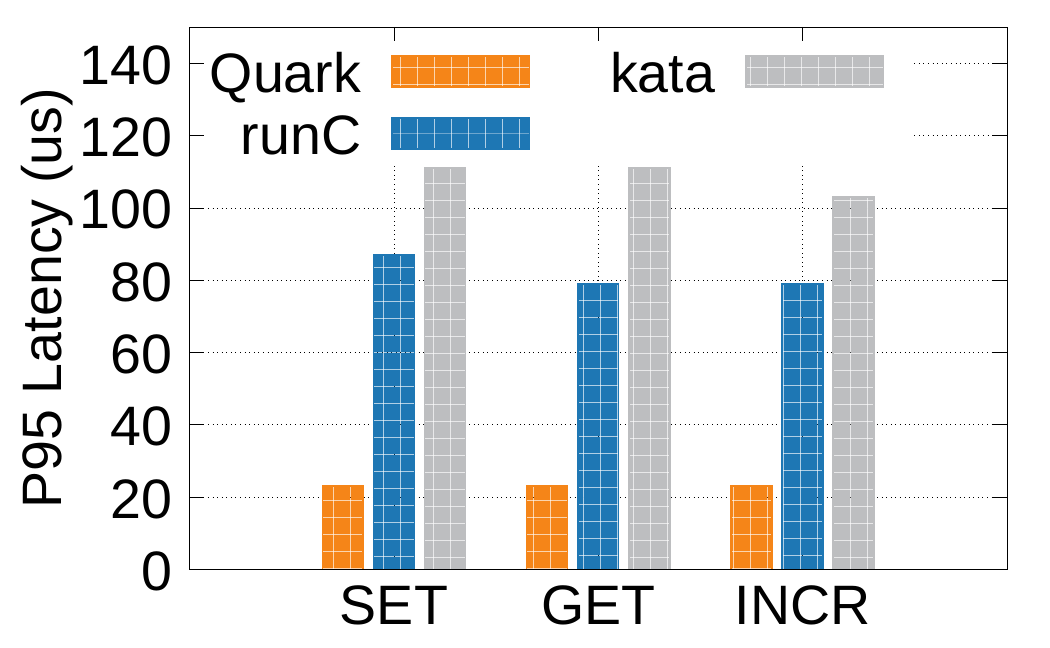}}
            \label{fig::redis_single_lat}
    	\end{minipage}}
    \subfigure[Single Connection]{
		\begin{minipage}[t]{0.3\linewidth}{
				\includegraphics[width=\linewidth]{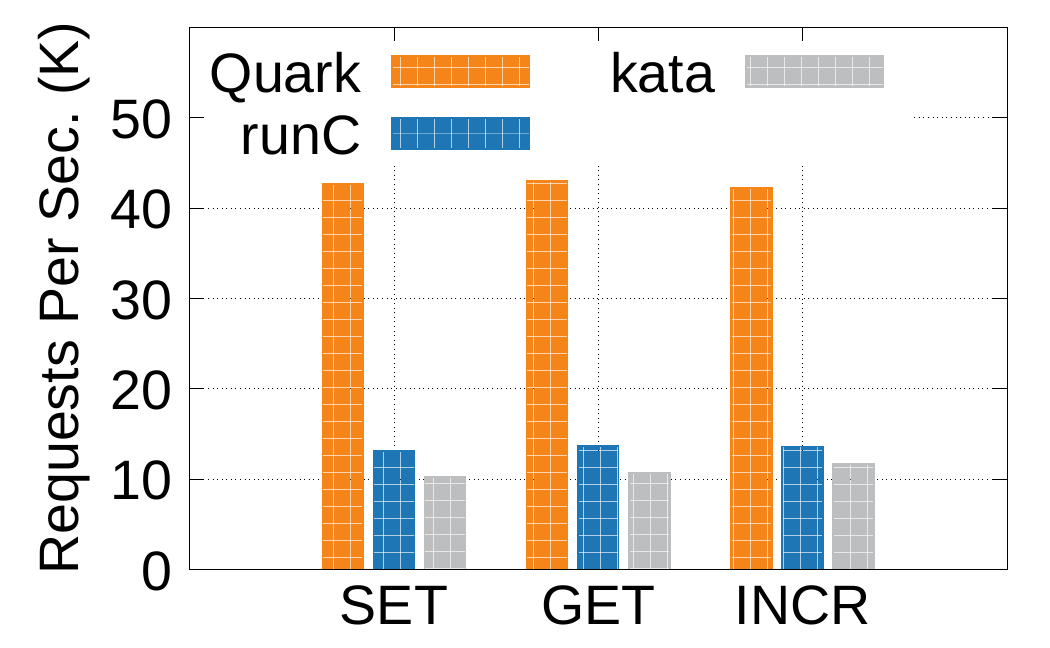}}
            \label{fig::redis_single_rps}
    	\end{minipage}}
            \subfigure[Multi-connection GET]{
		\begin{minipage}[t]{0.3\linewidth}{
				\includegraphics[width=\linewidth]{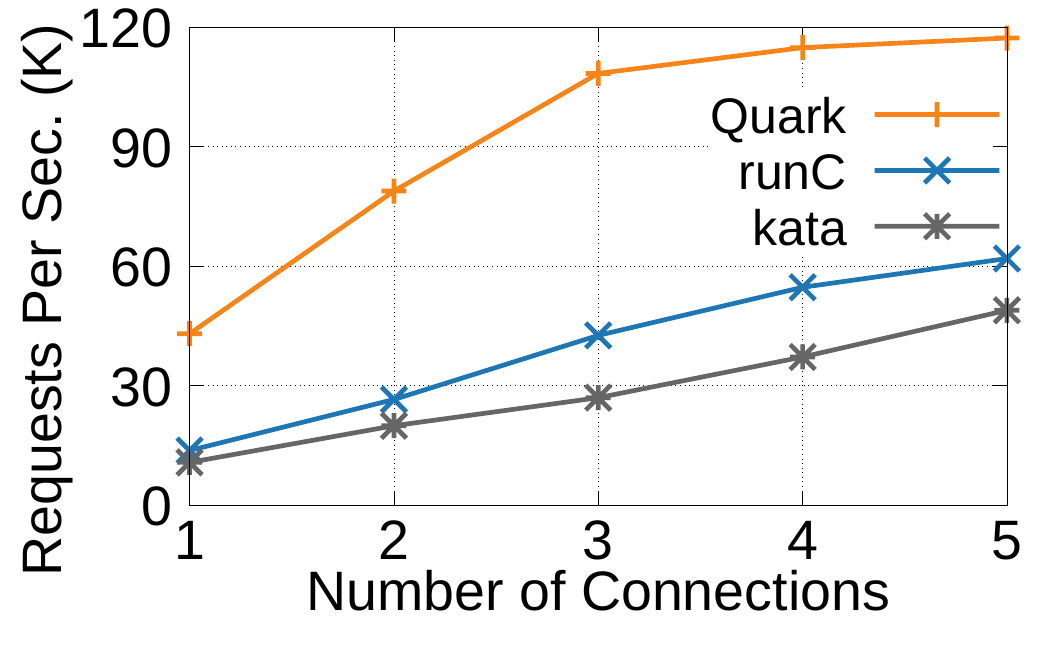}}
            \label{fig::redis_multi_get_rps}
    	\end{minipage}}
\caption{Redis Performance. Quark achieves lower response latency, higher throughput, and better scalability.}
\label{fig::redis}
\end{figure*}

\section{Implementaion}

\begin{table}[h]
    \footnotesize
    \centering
     \begin{tabular}{ c|c|c}
     \textbf{Component} & \textbf{Main Submodule}  & \textbf{$\sim$LOC}  \\
         \hline
        \multirow{6}{*}{QKernel} & Syscall Layer &  14.6 K \\
        \cline{2-3}
         & Memory Mananger &  5.7 K \\
  \cline{2-3}
        & File System &  34 K \\
    \cline{2-3}
        & Thread Manager &  9.2 K \\
    \cline{2-3}
      & Kernel Utility (e.g., timer) &  14.2 K \\
  \cline{2-3}
      & Network Socket &  13.2 K \\
      \hline
         QVisor & Runtime  &  28.4 K\\
         \hline
        \multirow{2}{*}{TSoR} & TSoR Service &  11.2 K\\
          \cline{2-3}
      & TSoR Client &  5 K\\
         \hline
        In Total &  &  135.5 K\\
        \end{tabular}
        \caption{Quark Codebase}
        \label{tab:code}
\end{table}

We developed Quark from the ground up using the Rust programming language, which features memory safety and high performance. Table \ref{tab:code} provides a detailed breakdown of the Lines of Code (LOC) across Quark's core subsystems. Notably, we have engineered the TSoR module as a separate component to facilitate sharing TSoR service among multiple sandboxes. Quark is available as an open-source project.

In the development of Quark, we place a strong emphasis on compatibility to seamlessly support container images without requiring any modifications to user code, recompilations, or dynamic pre-loading, which could introduce security concerns and additional burdens for our users in practical deployments. For container images targeting Linux, QKernel supports about 150 system calls. Notably, Quark fully supports TCP socket system calls while transparently improving the network performance with RDMA.

Quark is Kubernetes-ready and Docker-ready. Our implementation aims to be fully compatible with existing orchestration systems like Kubernetes and Docker. This ensures that Quark can be seamlessly integrated into existing Kubernetes clusters and take advantage of various tools within the Kubernetes ecosystem. To achieve Kubernetes compatibility, Quark adheres to the Container Runtime Interface (CRI) specification \cite{cri}, the API set that Kubernetes defines for interaction with container runtimes. 
For Docker compatibility, Quark complies with the Open Container Initiative's (OCI) Runtime Specification \cite{oci}. Furthermore, our TSoR module is fully aligned with the Kubernetes Networking Model, allowing it to be orchestrated by networking control planes such as the Kubernetes API Server.

\section{Evaluation}
\label{sec::evaluation}
\subsection{Setup}

\textbf{Testbed}. Our testbed comprises an RDMA-capable cluster with x86 servers connected to a 100Gbps Arista 716032-CQ switch. Each server is equipped with two Intel Xeon Gold 5218 processors, 96GB memory, and a 100Gbps dual-port Nvidia ConnectX-6 Dx NIC. We run Ubuntu 20.04 with kernel 5.15 and RoCEv2 for RDMA. We use Kubernetes v1.21 and Docker v20.10.

\textbf{Comparison Baselines}. We choose two common container runtimes: 1) \textit{runC} \cite{runc} is the default runtime for Docker and represents a widely-adopted, non-sandboxed container runtime. By directly operating as a host OS process without involving a guest kernel and VMM, runC generally outperforms VM-based secure container runtimes, making it a strong baseline for performance comparisons. For container networking, we configure runC with  Flannel \cite{Flannel},  one widely-used  Container Networking Interface (CNI) plugin. We use runC version 1.1.4 for experiments.  2) \textit{Kata} \cite{kata} is a secure container runtime that runs containers inside a lightweight virtual machine with its own Linux guest kernel and  QEMU as the VMM. Like runC, we configure Kata with Flannel to enable container networking.  We use Kata version 1.13. 

\subsection{End-to-end Application IO Performance}
\subsubsection{Redis}
Redis \cite{redis} is an in-memory key-value data store. We use the official Redis docker image (v7.0.5) and the built-in Redis-benchmark utility to collect the metrics of latency and throughput while running with different runtimes. We set up the server and client in different containers deployed on different hosts. We select SET, GET, and INCR operations where INCR means incrementing the number stored at \textit{key} by one. 

Figure \ref{fig::redis_single_lat} presents the latency for executing SET, GET, and INCR operations with the default data size of 3 bytes. To focus on measuring I/O performance and avoid internal queuing within the application, we set up a single connection for this experiment. For SET, Quark's  P95 latency is only 23 us, 79.3\% percent lower than kata. For all test operations, Quark achieves significantly lower latency than others. This is primarily attributable to Quark's RDMA-based TSoR networking and its efficient QCall mechanism for I/O operations.  

Figure \ref{fig::redis_single_rps} presents the throughput comparisons with the metric of Request Per Second (RPS) for SET, GET, and INCR operations. TSoR achieves higher than other solutions for all three operations. For SET operation, Quark achieves about 42K RPS, 4.1x times higher than kata, which highlights the effectiveness of both TSoR and QCall mechanisms.

Figure \ref{fig::redis_multi_get_rps} shows the throughput comparisons while increasing the number of connections. Quark almost linearly scales while varying the number of connections from 1 to 3. For GET operation, TSoR achieves 42K RPS for a single connection and 117K RPS for five connections, which is 2.43x higher than kata. While scaling to more connections, performance is bottlenecked by the queue buildup within Redis rather than the network.    

\begin{figure*}[htbp]
	\centering
    \subfigure[Node.js Latency]{
		\begin{minipage}[t]{0.3\linewidth}{
				\includegraphics[width=\linewidth]{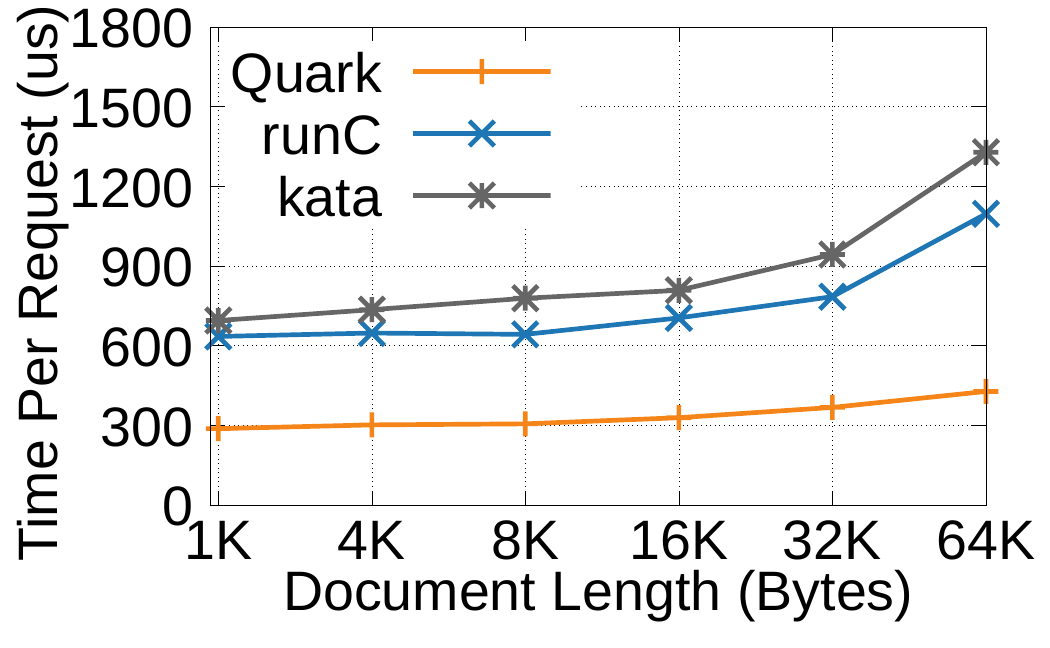}}
            \label{fig::nodejs_lat}
    	\end{minipage}}
    \subfigure[Node.js Throughput]{
		\begin{minipage}[t]{0.3\linewidth}{
				\includegraphics[width=\linewidth]{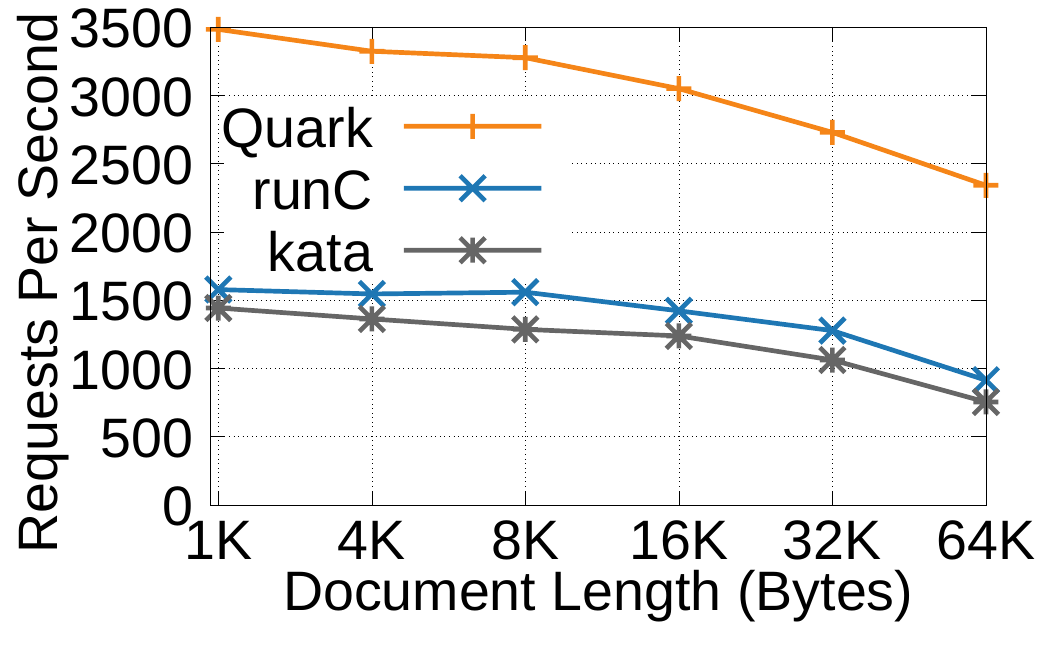}}
     \label{fig::nodejs_rps}
    	\end{minipage}}
    \subfigure[Etcd Throughput]{
		\begin{minipage}[t]{0.3\linewidth}{
				\includegraphics[width=\linewidth]{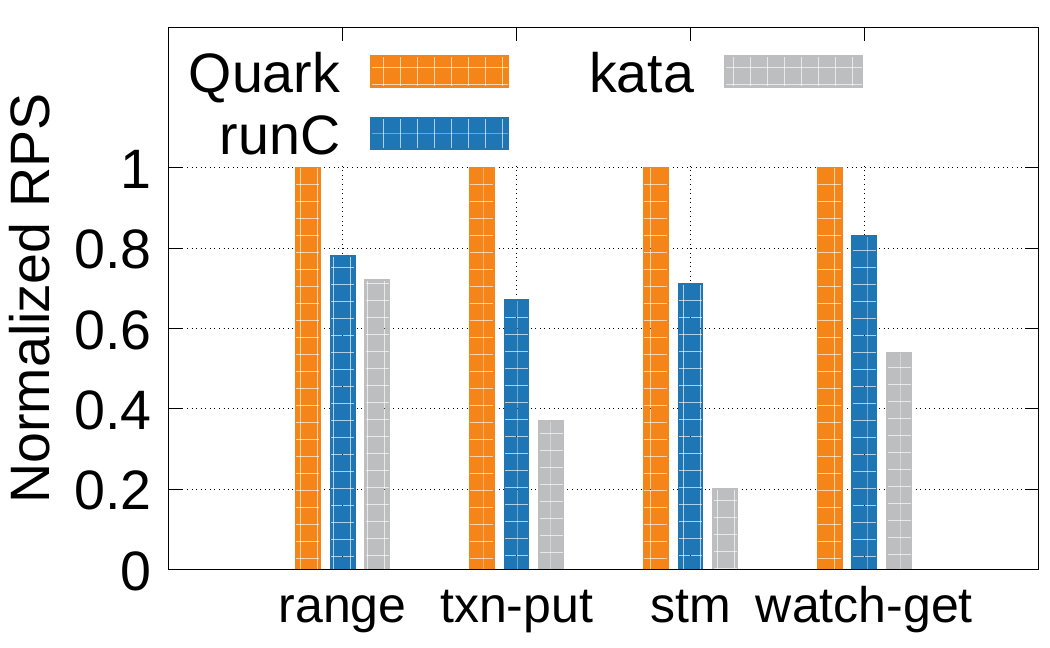}}
     \label{fig::etcd}
    	\end{minipage}}
\caption{Node.js and etcd performance. Quark achieves lower response latency and higher throughput.}
\end{figure*}

\subsubsection{Node.js}
Node.js \cite{nodejs} is a software platform for server-side networking applications that can act as a web server with support of HTTP and socket. We use node.js image to set up a web server and then use Apache HTTP server benchmarking tool (ab) \cite{ab} to generate requests from the client side. The server and client run on different host machines.

Figure \ref{fig::nodejs_lat} shows the response latency while varying the document length returned by the Node.js web server. Quark completes the request faster than others, especially when transferring documents with large sizes. For transferring a $64KB$ document, TSoR achieves 67.8\% lower latency than kata because the RDMA-based data path provides higher throughput with lower stack overhead. 

Figure \ref{fig::nodejs_rps} shows the throughput for transferring documents with different sizes. For a $64KB$ document, TSoR achieves 2.57x times higher throughput than runC, the performance upper bound for other solutions.

\subsubsection{Etcd}
Etcd \cite{etcd} is a distributed key-value store that uses the Raft to achieve strong consensus. We use the etcd image (v3.0.0) to set up a server and run the official benchmarking tool with ten connections. The server and client run over the different host machines. We select four common test cases for etcd:
 \textit{range} request is to get multiple keys. \textit{txn-put} is to write a single key within a transaction (txn). \textit{stm} is the implementation of software transaction memory. A \textit{watch-get} request tells etcd to notify the requester of getting to any provided keys.

Figure \ref{fig::etcd} shows the throughput as Request Per Second (RPS) while executing different benchmarking tests.   To unify the scale of different test cases, we normalize the RPS of other baselines to TSoR's performance numbers. For five common test cases, TSoR achieves higher RPS than other baselines. For example, for a typical test case of \textit{txn-put}, TSoR's RPS is 1.49x times higher than runC. 

\subsection{Network Microbenchmark}

\begin{figure}[ht!]
    \centering
	\subfigure[Throughput]{
		\begin{minipage}[t]{0.48\linewidth}{
				\includegraphics[width=\linewidth]{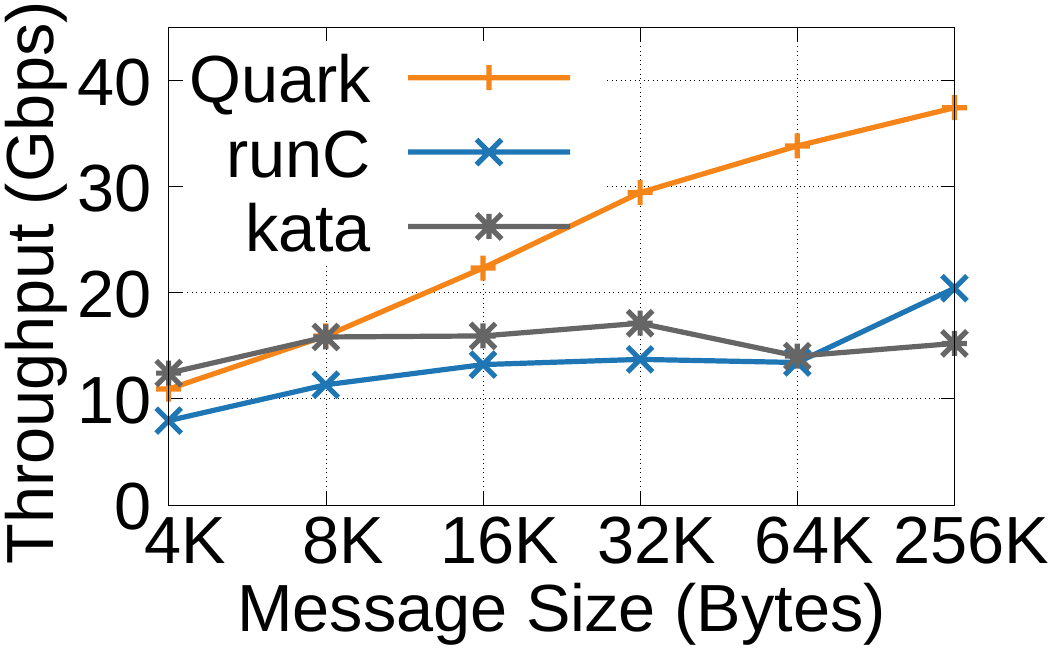}}
            \label{fig::single-throughput}
    	\end{minipage}}
        \subfigure[Latency]{
            \begin{minipage}[t]{0.48\linewidth}{
                    \includegraphics[width=\linewidth]{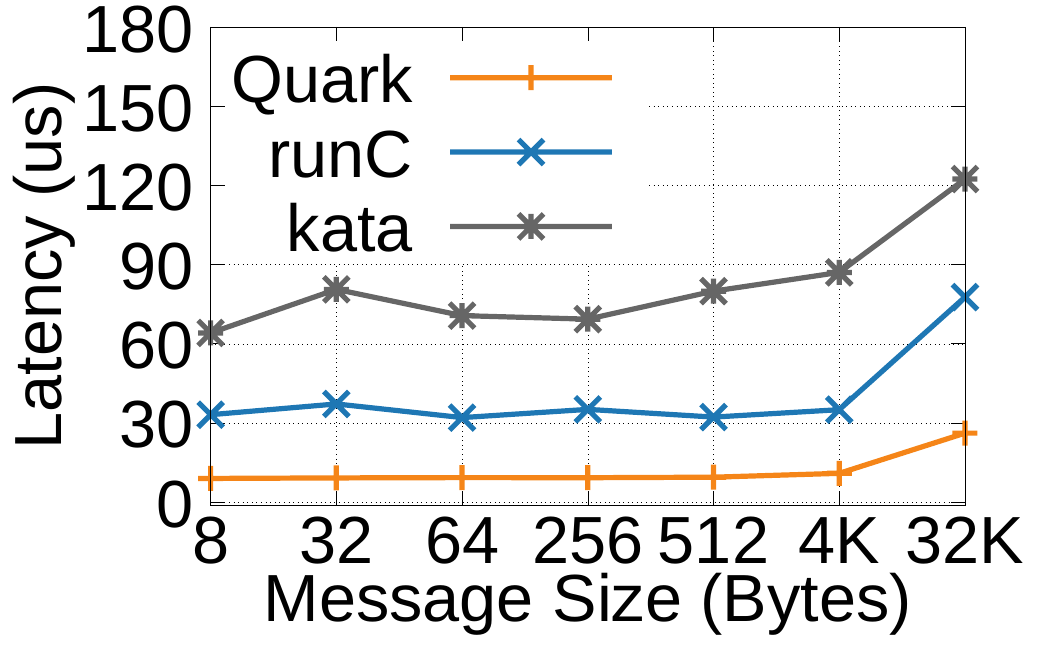}}
                \label{fig::single-latency}
            \end{minipage}}
\caption{TCP connection throughput and latency between a pair of containers across hosts with \textit{iperf3}.}
\end{figure}

\textbf{Throughput}: In Figure \ref{fig::single-throughput}, we present a comparison of throughput performance for a single TCP connection between paired containers across hosts. We use iPerf3 to generate TCP traffic while varying the message size. We focus on the throughput performance for transferring the message with a relatively large size (larger than $4KB$). Because in practice, most of the throughput-hungry scenarios like file transferring use relatively large message sizes. Varying message size from $4KB$ to $256KB$, Quark consistently outperforms other setups. The advantages of Quark's TSoR become even more pronounced as the message size increases. This is because TSoR's RDMA-based data path allows for greater maximum throughput compared to alternatives constrained by the kernel's TCP/IP stack.
For a typical message size of  $64KB$, TSoR achieves 34.3 Gbps throughput, which is even higher than runC, which is a non-sandboxed runtime that operates without the overhead of a guest kernel and VMM. It's worth noting that Quark, while employing a sandboxed model, offers stronger security isolation compared to runC. Quark increases the iperf throughout by 2.46x for the message size of  $256KB$ compared to kata.

\textbf{Latency}: Figure \ref{fig::single-latency} shows latency comparison for a cross-host TCP connection. We use NPtcp to measure the average latency for executing 1000 times message transferring with varying the message size. We focus on the latency performance for a small message which is common in typical microservices using RPCs. TSoR achieves constantly lower latency for small messages. For 64-byte messages, TSoR achieves 9.3 us latency which is 70.9\% lower than runC and 86.8\% lower than kata. Due to removing lots of layers of security and virtualization, runC generally plays as a performance ceiling for host OS TCP/IP stack-based container network solutions. By leveraging an RDMA-based data path, TSoR avoids the limitations imposed by the Linux kernel's TCP stack, achieving superior latency performance in comparison to runC.

\begin{table}[h]
    \small
    \centering
     \begin{tabular}{ c|c|c }
         \textbf{Quark-TSoR} & \textbf{RunC-Flannel}  & \textbf{Kata-Flannel}\\
        \hline
         157.55$_{\pm21}$ & 504.65$_{\pm103}$  & 834.6$_{\pm157}$\\
    
        \end{tabular}
        \caption{TCP Connection Setup Time ($\mu s _{\pm stddev}$). }
        \label{tab::tcp-setup}
\end{table}

\textbf{TCP Connection Establishment Time}. Table \ref{tab::tcp-setup} shows the TCP connection establishment time. Using TSoR, the establishment of a TCP connection takes a much shorter time. TSoR only needs less than half of the establishment time compared with runC. As we described before, although TSoR requires one RTT between server and client, similar to the handshake process of standard TCP, RDMA has a much lower RTT latency. So TSoR can benefit from RDMA to establish connections faster. We notice that TSoR's establishment time has much less variation compared with others. Because RDMA's low latency performance is more stable than the host OS TCP/IP stack. As the experiment shows,  TSoR can more efficiently handle connection establishment for many short-lived TCP connections, which is challenging for lots of prior work \cite{zhuo2019slim,li2019socksdirect}.  Particularly aligned with the short-term execution model of microservices and serverless computing, the short-lived TCP connection is a common scenario.

\subsection{Container Startup Performance}
\begin{figure}[htbp]
     \centering
\includegraphics[width=\linewidth]{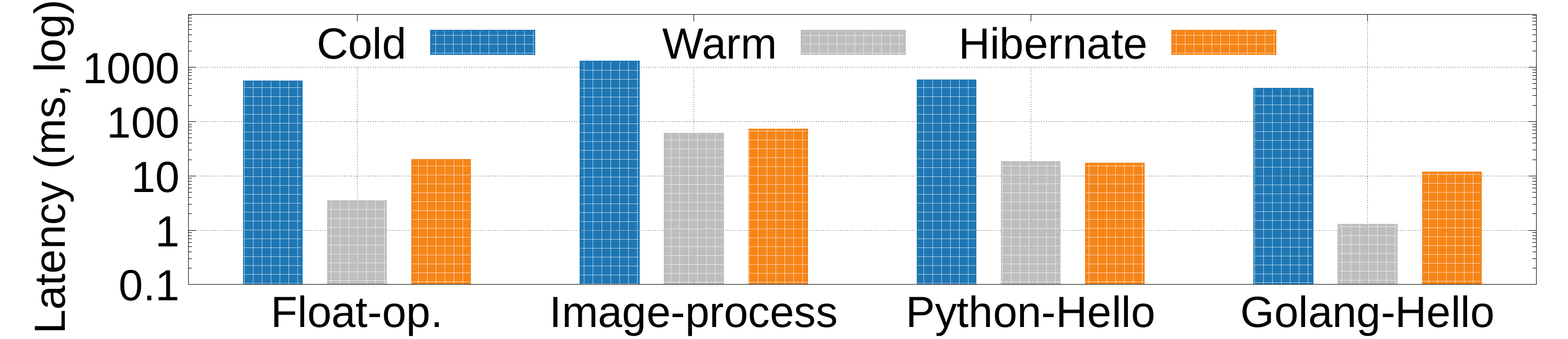}
 \caption{Response Latency for HTTP Requests.}
 \label{fig::hb_lat}
\end{figure}
\textbf{Request Response Latency}: Figure \ref{fig::hb_lat} shows the response latency for different modes of container startup upon the arrival of user requests. In this experiment, we set up four containerized services: Float Processing, Image Processing, Python Helloworld, and Golang Helloworld. All services are triggered via the HTTP requests from the client.  For \textit{Cold} mode, the container startup process includes setup runtime, application initialization, and application processing.  For \textit{Warm} mode, containers are alive before requests arrive. For \textit{Hibernation} mode, containers are in the hibernation mode, the customized mode enabled in Quark. We measure the end-to-end latency from the user side. From the results, we can see that the hibernation mode significantly reduces the response latency compared to the cold mode. Taking the Float-op as the example, the hibernation mode is 19.9 ms, 96.5\% lower than the cold mode. Also, the hibernation mode achieves a comparable latency with the warm mode. Note that the y-axis is log-scale. Additionally, these applications also demonstrate that Quark can benefit different language runtimes such as Python, Golang, and Java.

 \begin{figure}[htbp]
	\centering
                 \includegraphics[width=\linewidth]{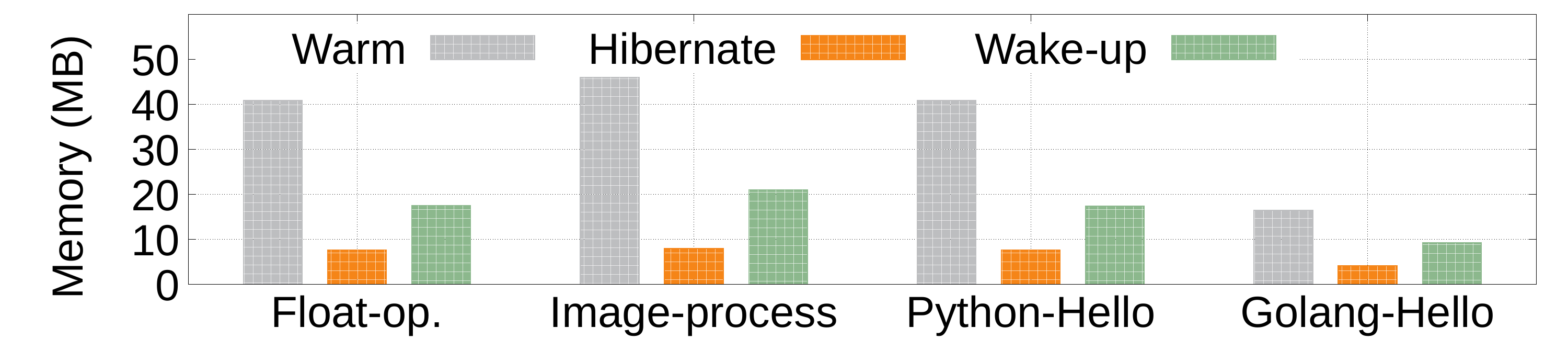}
        \caption{Idle Memory Cost for Container Start Modes.}
        \label{fig::hb_mem}
\end{figure}

\textbf{Idel Memory Cost:} Figure \ref{fig::hb_mem} shows the idle memory usage across different container startup modes, using the same applications as in the response latency experiment. Memory consumption is measured using the Linux utility \textit{pmap} to report the Proportional Set Size (PSS). In our test, we report the PSS data with ten concurrent running containers. In the cold mode, containers are terminated after request processing, resulting in zero idle memory cost.  
The \textit{wake-up} mode refers to containers resuming from hibernation by restoring application memory and processes. 
Unlike the warm mode, the wake-up mode employs on-demand memory loading while swapping in the application memory from disk. 
Our results indicate that hibernation mode significantly reduces idle memory consumption compared to the warm mode. For instance, Hibernation containers consume 81.3\% less memory than the warm mode. Even during active processing, the wake-up mode maintains a lower memory footprint due to its on-demand memory loading technique.





\subsection{Sandbox Memory Overhead}
\begin{table}[h]
    \small
    \centering
     \begin{tabular}{ c|c|c|c }
         Test &\textbf{Quark} & \textbf{runC}  & \textbf{Kata}\\
        \hline
       busybox memory usage (MB) &  11.8 & N/A & 184.3 \\
        \end{tabular}
        \caption{Memory Overhead of Sandboxes.}
        \label{tab:resourcesaving}
\end{table}

In this experiment, we evaluate the memory overhead associated with the sandbox, which includes both the guest kernel and VMM. We initiate a container using the commonly used \textit{busybox} image on both Quark and Kata, then measure the memory consumption of the guest kernel and VMM processes. It's worth noting that runC, being a non-sandboxed container runtime, incurs no such memory overhead. As Table \ref{tab:resourcesaving} shows, Quark's sandbox consumes considerably less memory compared to Kata's. This is primarily due to QKernel and QVisor being more lightweight than the Linux kernel and QEMU used by Kata, respectively.

\section{Conclusion}
In this work, we present Quark, a high-performance secure container runtime.  We've built it from scratch, combining a guest VM kernel called QKernel and a VMM called QVisor. 
The QKernel-QVisor codesign bring three major improvements: performant networking solution TSoR, fast container startup using Hibernation mode, and efficient syscall mechanism QCall. Comprehensive evaluations show that Quark is high-performance. Industy-standard and open-sourced codebases have great potential to prototype future innovation.  



\newpage
\bibliographystyle{unsrt}
\bibliography{references}

\begin{thebibliography}{10}

\bibitem{containerization}
{Containerization}.
\newblock \url{https://www.ibm.com/cloud/learn/containerization}, 2022.

\bibitem{mahgoub2022orion}
Ashraf Mahgoub, Edgardo~Barsallo Yi, Karthick Shankar, Sameh Elnikety, Somali Chaterji, and Saurabh Bagchi.
\newblock {ORION} and the three rights: Sizing, bundling, and prewarming for serverless {DAGs}.
\newblock In {\em 16th USENIX Symposium on Operating Systems Design and Implementation (OSDI 22)}, pages 303--320, 2022.

\bibitem{sartakov2022cap}
Vasily~A Sartakov, Llu{\'\i}s Vilanova, David Eyers, Takahiro Shinagawa, and Peter Pietzuch.
\newblock {CAP-VMs}:{Capability-Based} isolation and sharing in the cloud.
\newblock In {\em 16th USENIX Symposium on Operating Systems Design and Implementation (OSDI 22)}, pages 597--612, 2022.

\bibitem{qiu2020firm}
Haoran Qiu, Subho~S Banerjee, Saurabh Jha, Zbigniew~T Kalbarczyk, and Ravishankar~K Iyer.
\newblock {FIRM}: An intelligent fine-grained resource management framework for {SLO-Oriented} microservices.
\newblock In {\em 14th USENIX Symposium on Operating Systems Design and Implementation (OSDI 20)}, pages 805--825, 2020.

\bibitem{van2022blackbox}
Alexander Van't~Hof and Jason Nieh.
\newblock {BlackBox}: A container security monitor for protecting containers on untrusted operating systems.
\newblock In {\em 16th USENIX Symposium on Operating Systems Design and Implementation (OSDI 22)}, pages 683--700, 2022.

\bibitem{Groundhog}
Mohamed Alzayat, Jonathan Mace, Peter Druschel, and Deepak Garg.
\newblock Groundhog: Efficient request isolation in faas, 2022.

\bibitem{liu2023doing}
David~H Liu, Amit Levy, Shadi Noghabi, and Sebastian Burckhardt.
\newblock Doing more with less: Orchestrating serverless applications without an orchestrator.
\newblock In {\em 20th USENIX Symposium on Networked Systems Design and Implementation (NSDI 23)}, pages 1505--1519, 2023.

\bibitem{xu2023dirigo}
Le~Xu, Divyanshu Saxena, Neeraja~J Yadwadkar, Aditya Akella, and Indranil Gupta.
\newblock Dirigo: Self-scaling stateful actors for serverless real-time data processing.
\newblock {\em arXiv preprint arXiv:2308.03615}, 2023.

\bibitem{gVisor}
The gVisor Authors.
\newblock {gVisor}.
\newblock \url{https://gvisor.dev/}, 2022.

\bibitem{firecracker}
Alexandru Agache, Marc Brooker, Alexandra Iordache, Anthony Liguori, Rolf Neugebauer, Phil Piwonka, and Diana-Maria Popa.
\newblock Firecracker: Lightweight virtualization for serverless applications.
\newblock In {\em 17th USENIX symposium on networked systems design and implementation (NSDI 20)}, pages 419--434, 2020.

\bibitem{RunD}
Zijun Li, Jiagan Cheng, Quan Chen, Eryu Guan, Zizheng Bian, Yi~Tao, Bin Zha, Qiang Wang, Weidong Han, and Minyi Guo.
\newblock {RunD}: A lightweight secure container runtime for high-density deployment and high-concurrency startup in serverless computing.
\newblock In {\em 2022 USENIX Annual Technical Conference (USENIX ATC 22)}, pages 53--68, 2022.

\bibitem{mspodsandbox}
{Pod Sandboxing with Azure Kubernetes Service (AKS)}.
\newblock \url{https://learn.microsoft.com/en-us/azure/aks/use-pod-sandboxing}, 2023.

\bibitem{cve2017}
{CVE-2017-5123}.
\newblock \url{https://nvd.nist.gov/vuln/detail/CVE-2017-5123}, 2023.

\bibitem{cve2019}
{CVE-2019-5736 Detail}.
\newblock \url{https://nvd.nist.gov/vuln/detail/CVE-2019-5736}, 2023.

\bibitem{bellard2005qemu}
Fabrice Bellard.
\newblock Qemu, a fast and portable dynamic translator.
\newblock In {\em USENIX annual technical conference, FREENIX Track}, volume~41, page~46. Califor-nia, USA, 2005.

\bibitem{kvm}
{Kernel Virtual Machine}.
\newblock \url{https://www.linux-kvm.org/page/Main_Page}, 2022.

\bibitem{kata}
{Kata Containers}.
\newblock \url{https://katacontainers.io/}, 2022.

\bibitem{zhang2021demikernel}
Irene Zhang, Amanda Raybuck, Pratyush Patel, Kirk Olynyk, Jacob Nelson, Omar S~Navarro Leija, Ashlie Martinez, Jing Liu, Anna~Kornfeld Simpson, Sujay Jayakar, et~al.
\newblock The demikernel datapath os architecture for microsecond-scale datacenter systems.
\newblock In {\em Proceedings of the ACM SIGOPS 28th Symposium on Operating Systems Principles}, pages 195--211, 2021.

\bibitem{zhang2022justitia}
Yiwen Zhang, Yue Tan, Brent Stephens, and Mosharaf Chowdhury.
\newblock Justitia: Software {Multi-Tenancy} in hardware {Kernel-Bypass} networks.
\newblock In {\em 19th USENIX Symposium on Networked Systems Design and Implementation (NSDI 22)}, pages 1307--1326, 2022.

\bibitem{marty2019snap}
Michael Marty, Marc de~Kruijf, Jacob Adriaens, Christopher Alfeld, Sean Bauer, Carlo Contavalli, Michael Dalton, Nandita Dukkipati, William~C Evans, Steve Gribble, et~al.
\newblock Snap: A microkernel approach to host networking.
\newblock In {\em Proceedings of the 27th ACM Symposium on Operating Systems Principles}, pages 399--413, 2019.

\bibitem{guo2016rdma}
Chuanxiong Guo, Haitao Wu, Zhong Deng, Gaurav Soni, Jianxi Ye, Jitu Padhye, and Marina Lipshteyn.
\newblock Rdma over commodity ethernet at scale.
\newblock In {\em Proceedings of the 2016 ACM SIGCOMM Conference}, pages 202--215, 2016.

\bibitem{dalton2018andromeda}
Michael Dalton, David Schultz, Jacob Adriaens, Ahsan Arefin, Anshuman Gupta, Brian Fahs, Dima Rubinstein, Enrique~Cauich Zermeno, Erik Rubow, James~Alexander Docauer, et~al.
\newblock Andromeda: Performance, isolation, and velocity at scale in cloud network virtualization.
\newblock In {\em 15th USENIX symposium on networked systems design and implementation (NSDI 18)}, pages 373--387, 2018.

\bibitem{zhuo2019slim}
Danyang Zhuo, Kaiyuan Zhang, Yibo Zhu, Hongqiang~Harry Liu, Matthew Rockett, Arvind Krishnamurthy, and Thomas Anderson.
\newblock Slim:{OS} kernel support for a {Low-Overhead} container overlay network.
\newblock In {\em 16th USENIX Symposium on Networked Systems Design and Implementation (NSDI 19)}, pages 331--344, 2019.

\bibitem{ma2022survey}
Shaonan Ma, Teng Ma, Kang Chen, and Yongwei Wu.
\newblock A survey of storage systems in the rdma era.
\newblock {\em IEEE Transactions on Parallel and Distributed Systems}, 2022.

\bibitem{su2022pipedevice}
Qiang Su, Chuanwen Wang, Zhixiong Niu, Ran Shu, Peng Cheng, Yongqiang Xiong, Dongsu Han, Chun~Jason Xue, and Hong Xu.
\newblock Pipedevice: a hardware-software co-design approach to intra-host container communication.
\newblock In {\em Proceedings of the 18th International Conference on emerging Networking EXperiments and Technologies}, pages 126--139, 2022.

\bibitem{du2020catalyzer}
Dong Du, Tianyi Yu, Yubin Xia, Binyu Zang, Guanglu Yan, Chenggang Qin, Qixuan Wu, and Haibo Chen.
\newblock Catalyzer: Sub-millisecond startup for serverless computing with initialization-less booting.
\newblock In {\em Proceedings of the Twenty-Fifth International Conference on Architectural Support for Programming Languages and Operating Systems}, pages 467--481, 2020.

\bibitem{yu2020characterizing}
Tianyi Yu, Qingyuan Liu, Dong Du, Yubin Xia, Binyu Zang, Ziqian Lu, Pingchao Yang, Chenggang Qin, and Haibo Chen.
\newblock Characterizing serverless platforms with serverlessbench.
\newblock In {\em Proceedings of the 11th ACM Symposium on Cloud Computing}, pages 30--44, 2020.

\bibitem{ustiugov2021benchmarking}
Dmitrii Ustiugov, Plamen Petrov, Marios Kogias, Edouard Bugnion, and Boris Grot.
\newblock Benchmarking, analysis, and optimization of serverless function snapshots.
\newblock In {\em Proceedings of the 26th ACM International Conference on Architectural Support for Programming Languages and Operating Systems}, pages 559--572, 2021.

\bibitem{redis}
{Redis}.
\newblock \url{https://redis.io/}, 2022.

\bibitem{nodejs}
{Nodejs}.
\newblock \url{https://nodejs.org/en/}, 2022.

\bibitem{etcd}
{etcd}.
\newblock \url{https://etcd.io/}, 2022.

\bibitem{iperf}
{Iperf3}.
\newblock \url{https://iperf.fr/}, 2022.

\bibitem{runc}
{runC Containers}.
\newblock \url{https://github.com/opencontainers/runc}, 2022.

\bibitem{kuenzer2021unikraft}
Simon Kuenzer, Vlad-Andrei B{\u{a}}doiu, Hugo Lefeuvre, Sharan Santhanam, Alexander Jung, Gaulthier Gain, Cyril Soldani, Costin Lupu, {\c{S}}tefan Teodorescu, Costi R{\u{a}}ducanu, et~al.
\newblock Unikraft: fast, specialized unikernels the easy way.
\newblock In {\em Proceedings of the Sixteenth European Conference on Computer Systems}, pages 376--394, 2021.

\bibitem{Flannel}
{Flannel}.
\newblock \url{https://github.com/coreos/flannel/}, 2022.

\bibitem{Weave}
{Weave Net}.
\newblock \url{https://github.com/weaveworks/weave}, 2022.

\bibitem{Cilium}
{Cilium}.
\newblock \url{https://cilium.io/}, 2022.

\bibitem{wei2023no}
Xingda Wei, Fangming Lu, Tianxia Wang, Jinyu Gu, Yuhan Yang, Rong Chen, and Haibo Chen.
\newblock No provisioned concurrency: Fast {RDMA-codesigned} remote fork for serverless computing.
\newblock In {\em 17th USENIX Symposium on Operating Systems Design and Implementation (OSDI 23)}, pages 497--517, 2023.

\bibitem{wei2022krcore}
Xingda Wei, Fangming Lu, Rong Chen, and Haibo Chen.
\newblock {KRCORE}: A microsecond-scale {RDMA} control plane for elastic computing.
\newblock In {\em 2022 USENIX Annual Technical Conference (USENIX ATC 22)}, pages 121--136, 2022.

\bibitem{prakash2022portkey}
Chandra Prakash, Debadatta Mishra, Purushottam Kulkarni, and Umesh Bellur.
\newblock Portkey: Hypervisor-assisted container migration in nested cloud environments.
\newblock In {\em Proceedings of the 18th ACM SIGPLAN/SIGOPS International Conference on Virtual Execution Environments}, pages 3--17, 2022.

\bibitem{humphries2021case}
Jack~Tigar Humphries, Kostis Kaffes, David Mazi{\`e}res, and Christos Kozyrakis.
\newblock A case against (most) context switches.
\newblock In {\em Proceedings of the Workshop on Hot Topics in Operating Systems}, pages 17--25, 2021.

\bibitem{zhou2023userspace}
Zhe Zhou, Yanxiang Bi, Junpeng Wan, Yangfan Zhou, and Zhou Li.
\newblock Userspace bypass: Accelerating syscall-intensive applications.
\newblock In {\em 17th USENIX Symposium on Operating Systems Design and Implementation (OSDI 23)}, pages 33--49, 2023.

\bibitem{vmexit}
Zeyu Mi, Dingji Li, Haibo Chen, Binyu Zang, and Haibing Guan.
\newblock (mostly) exitless {VM} protection from untrusted hypervisor through disaggregated nested virtualization.
\newblock In {\em 29th USENIX Security Symposium (USENIX Security 20)}, pages 1695--1712. USENIX Association, August 2020.

\bibitem{reap}
Dmitrii Ustiugov, Plamen Petrov, Marios Kogias, Edouard Bugnion, and Boris Grot.
\newblock Benchmarking, analysis, and optimization of serverless function snapshots.
\newblock In {\em Proceedings of the 26th ACM International Conference on Architectural Support for Programming Languages and Operating Systems}, pages 559--572, 2021.

\bibitem{grpc}
{gRPC}.
\newblock \url{https://grpc.io/}, 2022.

\bibitem{amqparch}
{AMQP Architecture}.
\newblock \url{https://www.amqp.org/product/architecture}, 2022.

\bibitem{niu2017network}
Zhixiong Niu, Hong Xu, Dongsu Han, Peng Cheng, Yongqiang Xiong, Guo Chen, and Keith Winstein.
\newblock Network stack as a service in the cloud.
\newblock In {\em Proceedings of the 16th ACM Workshop on Hot Topics in Networks}, pages 65--71, 2017.

\bibitem{niu2021netkernel}
Zhixiong Niu, Qiang Su, Peng Cheng, Yongqiang Xiong, Dongsu Han, Keith Winstein, Chun~Jason Xue, and Hong Xu.
\newblock Netkernel: Making network stack part of the virtualized infrastructure.
\newblock {\em IEEE/ACM Transactions on Networking}, 30(3):999--1013, 2021.

\bibitem{bastion}
Jaehyun Nam, Seungsoo Lee, Hyunmin Seo, Phil Porras, Vinod Yegneswaran, and Seungwon Shin.
\newblock {BASTION}: A security enforcement network stack for container networks.
\newblock In {\em 2020 USENIX Annual Technical Conference (USENIX ATC 20)}, pages 81--95. USENIX Association, July 2020.

\bibitem{rdmaprogamming}
{RDMA Aware Networks Programming User Manual}.
\newblock \url{https://docs.nvidia.com/networking/pages/viewpage.action?pageId=34256548}, 2022.

\bibitem{kalia2016design}
Anuj Kalia, Michael Kaminsky, and David~G Andersen.
\newblock Design guidelines for high performance {RDMA} systems.
\newblock In {\em 2016 USENIX Annual Technical Conference (USENIX ATC 16)}, pages 437--450, 2016.

\bibitem{kong2022collie}
Xinhao Kong, Yibo Zhu, Huaping Zhou, Zhuo Jiang, Jianxi Ye, Chuanxiong Guo, and Danyang Zhuo.
\newblock Collie: Finding performance anomalies in {RDMA} subsystems.
\newblock In {\em 19th USENIX Symposium on Networked Systems Design and Implementation (NSDI 22)}, pages 287--305, 2022.

\bibitem{wang2019vsocket}
Dongyang Wang, Binzhang Fu, Gang Lu, Kun Tan, and Bei Hua.
\newblock vsocket: virtual socket interface for rdma in public clouds.
\newblock In {\em Proceedings of the 15th ACM SIGPLAN/SIGOPS International Conference on Virtual Execution Environments}, pages 179--192, 2019.

\bibitem{bonwick1994slab}
Jeff Bonwick et~al.
\newblock The slab allocator: An object-caching kernel memory allocator.
\newblock In {\em USENIX summer}, volume~16. Boston, MA, USA, 1994.

\bibitem{knowlton1965fast}
Kenneth~C Knowlton.
\newblock A fast storage allocator.
\newblock {\em Communications of the ACM}, 8(10):623--624, 1965.

\bibitem{iouring}
{io\_uring}.
\newblock \url{https://kernel.dk/io_uring.pdf}, 2023.

\bibitem{cri}
{ Container Runtime Interface (CRI)}.
\newblock \url{https://kubernetes.io/docs/concepts/architecture/cri/}, 2022.

\bibitem{oci}
{Open Container Initiative}.
\newblock \url{https://opencontainers.org/}, 2022.

\bibitem{ab}
{ Apache HTTP server benchmarking tool}.
\newblock \url{https://httpd.apache.org/docs/2.4/programs/ab.html}, 2022.

\bibitem{li2019socksdirect}
Bojie Li, Tianyi Cui, Zibo Wang, Wei Bai, and Lintao Zhang.
\newblock Socksdirect: Datacenter sockets can be fast and compatible.
\newblock In {\em Proceedings of the ACM Special Interest Group on Data Communication}, pages 90--103, 2019.

\end{thebibliography}


\end{document}